\documentclass[acmsmall,screen]{acmart} 
\usepackage{cleveref}
\usepackage{bm} 
\crefname{figure}{Fig.}{Figs.}
\Crefname{figure}{Figure}{Figures}

\AtBeginDocument{%
  }

\setcopyright{cc}
\setcctype{by}
\acmJournal{PACMHCI}
\acmYear{2026} \acmVolume{10} \acmNumber{8} \acmArticle{ISS011}
\acmMonth{11} \acmDOI{10.1145/3859323}

\usepackage{enumitem} 

\begin{document}

\title{Fluidic Body: Swarm Robots as a Fluid Extension of the Hand}


\author{Michikuni Eguchi}
\orcid{0009-0008-8370-3268}
\affiliation{%
 \institution{University of Tsukuba}
 \city{Ibaraki}
 \country{Japan}
}
\affiliation{%
  \institution{Cluster Metaverse Lab}
  \city{Tokyo}
  \country{Japan}
}
\email{eguchi@mvml.slis.tsukuba.ac.jp}

\author{Shigeo Yoshida}
\orcid{0000-0001-5381-0187}
\affiliation{%
  \institution{OMRON SINIC X Corporation}
  \city{Tokyo}
  \country{Japan}}
\email{shigeo.yoshida@sinicx.com}

\author{Mai Nishimura}
\orcid{0000-0003-2859-988X}
\affiliation{%
  \institution{OMRON SINIC X Corporation}
  \city{Tokyo}
  \country{Japan}}
\email{mai.nishimura@sinicx.com}

\author{Takefumi Hiraki}
\orcid{0000-0002-5767-3607}
\affiliation{%
 \institution{University of Tsukuba}
 \city{Ibaraki}
 \country{Japan}
}
\affiliation{%
  \institution{Cluster Metaverse Lab}
  \city{Tokyo}
  \country{Japan}
}
\email{hiraki@slis.tsukuba.ac.jp}

\renewcommand{\shortauthors}{Eguchi et al.}

\begin{abstract}
Swarm robots have attracted growing interest as a physical interface for human--computer interaction, since a swarm can reshape its formation to fit the objects and surroundings it works with.
For a human operator, however, adapting the swarm's behavior remains difficult. Most existing methods require the user to specify where the swarm should go, and they offer no intuitive way to change the collective's behavior as each task demands.
We present Fluidic Body, a system that lets a person operate a swarm of fluid-imitating robots as an extension of the hand. The system places a virtual fluid between the hand and the swarm, and the user adapts the swarm's behavior to each task by changing the fluid's properties through simple hand movements.
When the user opens the hand, the swarm becomes soft enough to deform and flow around obstacles. Closing the hand turns it into a hard, cohesive body that can push objects.
We evaluated the controller's performance and its operability in two simulation experiments and one demonstration with physical robots. First, our controller handled both obstacle avoidance and object transportation better than an existing method that behaves the same way in every task. Second, in a VR user study, 17 participants completed the combined avoidance-and-transportation task faster and more reliably than with that method. They also reported a lower workload and a stronger sense of agency. Finally, this experience carried over to physical robots in a real-world demonstration (88 participants).
These results suggest that treating a swarm as a fluid, a material people already understand, allows a robot swarm to work as a deformable extension of the human body.
\end{abstract}


\begin{teaserfigure}
  \includegraphics[width=\textwidth]{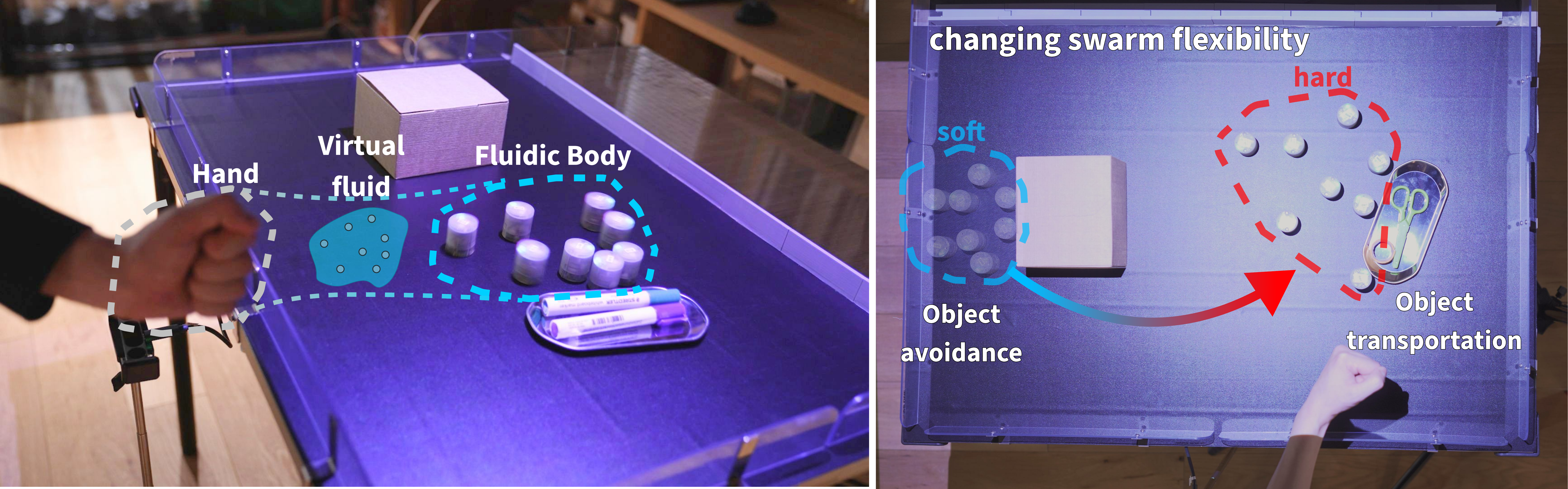}
  \caption{
  Fluidic Body lets a person operate a swarm of fluid-imitating robots as an extension of the hand.
  Left: the hand defines a region of virtual fluid, and the robots fill it and arrange themselves into a fluid body on the table.
  Right: the user changes the fluid's properties through hand movements to adapt the swarm's behavior to the task. For obstacle avoidance, a soft formation flows around the obstacles. For object transportation, the swarm hardens into a cohesive formation and pushes the object.
  }
  \Description{
  Two photographs of the tabletop system. Left: a person's hand held above a table, with an overlaid outline of a virtual fluid beside the hand and a group of small cylindrical robots outlined as a fluid body, among objects such as markers and a box. Right, titled ``Changing swarm flexibility'': the same robots in a loose soft formation (blue outline) flowing around a box for object avoidance, and in a tight hard formation (red outline) pushing an object for object transportation.
  }
  \label{fig:teaser}
\end{teaserfigure}


\maketitle

\section{Introduction}

Swarm robotics studies how many robots can act as one, producing collective behavior beyond the reach of any single robot~\cite{Sahin2005-ik, Brambilla2013-bg, Bonabeau1999-nv}.
A swarm can form patterns and assemble into shapes~\cite{5980269, Rubenstein2014-dm}.
It can also move through cluttered spaces as a group~\cite{Alonso-Mora2016-ka} and transport objects larger than its individual robots through cooperative pushing~\cite{10080994, Igarashi2010-if}.
The range of such tasks grows with the number of robots, and platforms of a thousand units have coordinated at that scale~\cite{Rubenstein2012-ry}.
Because a swarm can flexibly adapt to different tasks by reshaping its formation, it has drawn attention as a promising interface for human--computer interaction (HCI).
In HCI applications such as teleoperation~\cite{holobots, 10.1145/3613904.3642870} and tangible user interfaces~\cite{Le_Goc2016-dg}, a swarm works among people and everyday objects, and the situations it faces vary widely.

For swarm robots to serve a human operator, the missing piece is no longer the swarm controller, which computes how the robots move, but the interface through which a person directs them.
Methods such as potential fields~\cite{Khatib1986-si}, formation planning~\cite{Alonso-Mora2016-ka}, and reinforcement learning~\cite{Sui2021-zl} already generate diverse collective behaviors automatically.
What remains is an interface that lets a person reshape that behavior during operation to match the demands of each task.
Research on human-swarm interaction has explored directing the swarm through gestures and embodying it as an extension of the body~\cite{Le_Goc2016-dg, holobots, 10.1145/3613904.3642870}.
These methods, however, largely amount to specifying where the swarm should go, and changing how it behaves to fit each task is still unaddressed.
As a result, a person cannot yet fully draw out the potential of a swarm.


We propose \textit{Fluidic Body}, a system that lets a person operate a swarm of fluid-imitating robots as an extension of the hand (\cref{fig:teaser}).
To let a person shape the swarm's behavior itself, not only where it goes, we place a virtual fluid between the hand and the swarm.
The person changes the fluid's properties through hand movements, and the swarm produces the matching behavior as if it were that fluid.
Because such fluids are part of everyday life, a person already knows how they behave and can anticipate how the swarm will respond~\cite{battaglia2013physics, jacob2008rbi}.
An open hand softens the swarm into a deformable body that flows around obstacles.
Conversely, when the hand closes, the robots pack together into a hard mass that can push objects.
Through these simple hand movements alone, a person adapts the swarm's behavior to representative swarm tasks such as obstacle avoidance and object transportation~\cite{Alonso-Mora2016-ka, 10080994, Igarashi2010-if}.

Fluidic Body builds on smoothed particle hydrodynamics (SPH), which represents a fluid as a set of particles~\cite{Gingold1977-po}.
Treating each particle as a robot turns this fluid simulation into a controller that makes the swarm imitate fluid behavior~\cite{Pac2007-op}.
With this controller, the robots fill the region marked out by the hand and arrange themselves, and the person never has to address a robot individually.
We build on our previous SPH controller~\cite{Eguchi18062025, 10.1145/3721257.3734028} and map the hand's movements to the properties of the simulated fluid.
This mapping lets simple hand movements produce a range of fluid-imitating swarm behaviors.

We evaluate Fluidic Body along two axes, the controller's performance and the operability it gives a person.
For performance, we run a technical evaluation in simulation without a human operator.
It compares our controller against a fixed baseline that keeps the same behavior for every task, to test whether changing the fluid's flexibility helps across tasks.
Because no single behavior suits both tasks, the fixed baseline handles only one of them, whereas our controller adjusts the fluid to each task and handles both.
For operability, we run a VR user study and an in-the-wild demonstration in which participants operate the swarm themselves.
In the VR user study, 17 participants completed the combined avoidance-and-transportation task faster and more reliably than with the baseline and reported a lower workload and a stronger sense of agency.
In the in-the-wild demonstration, 88 participants operated the physical robots and agreed most strongly that their own hand movements drove the swarm.
This suggests that the sense of agency from the VR study carries over to real hardware.

Our contributions are as follows:
\begin{itemize}[leftmargin=4mm]
\item We propose Fluidic Body, a control interface that exposes a swarm's flexibility as a fluid property a person changes by hand to adapt the swarm's behavior to each task.
\item At the core of this interface, we design a mapping from three features of the hand to the fluid's viscosity and density and, through them, to the swarm.
The hand's shape switches the fluid between soft and hard settings, and its height adjusts how far the swarm spreads.
Its outline defines the region the swarm fills.
\item We conducted a three-part evaluation: a simulation comparison against a fixed baseline, a VR user study ($N=17$), and an in-the-wild demonstration with physical robots ($N=88$).
Adjusting the fluid adapted the swarm to each task and let participants operate it more efficiently than with the baseline, with a lower workload and a stronger sense of agency.
The in-the-wild demonstration suggested that this sense of agency carries over to physical robots.
\end{itemize}

\section{Related Work}

\subsection{Swarm User Interfaces}
Swarm user interfaces treat many small robots as a collective, tangible medium for human--computer interaction.
One line of applications uses the swarm as a display.
Zooids used tabletop robots as physical pixels~\cite{Le_Goc2016-dg}, and shape-changing units reconfigured their own form~\cite{Suzuki2019-ey}.
Kilobots demonstrated such swarm displays at the scale of a thousand robots~\cite{Rubenstein2012-ry, Rubenstein2014-dm}.
Other work extended these displays beyond the tabletop surface by coordinating tabletop robots with projected imagery~\cite{Hiraki2018-gi} and by using flying robots~\cite{Ghandeharizadeh2022-dv, Ghandeharizadeh2022-ii}.

Another line treats the robots as elements through which a person can act on the physical world.
Robots used as tangible interfaces let users manipulate elements directly on a tabletop~\cite{10.1145/1978942.1979384, Sugimoto2011-vi}, and a swarm moved objects to their destinations by pushing them under a person's direction~\cite{Igarashi2010-if}.
Such operation also extends beyond the person's own table, as mixed-reality systems let people at different sites co-manipulate a shared swarm for remote collaboration~\cite{holobots, Hiraki2018-dj, 10161412}.

Fluidic Body mainly targets this second line of applications, in which a person operates the swarm.
It uses a fluid-imitating swarm to make such physical interaction more efficient.

\subsection{Swarm Controller}
The robotics community has long studied how a swarm should be controlled~\cite{Bonabeau1999-nv, Sahin2005-ik, Brambilla2013-bg}.
Most of this work fixes the motion rule and its parameters in advance so that the swarm reaches a given goal on its own.
A representative example is reciprocal velocity obstacles (RVO), which prevent collisions by choosing each robot's velocity based on the motion of its neighbors and obstacles~\cite{Van_den_Berg2008-we}.
Later variants improved smoothness and extended the idea to aerial robots~\cite{Snape2010-sx, Gao2022-ur}.
Potential-field methods, another major approach, steer robots by attraction to goals and repulsion from obstacles~\cite{Khatib1986-si, Arkin1989-uy, Howard2002-vr}.
Whereas these methods act on each robot locally, formation planners treat the group as a whole and plan a formation that fits through cluttered and dynamic environments~\cite{Alonso-Mora2016-ka, Alonso-Mora2019-ti, Sabattini2017-od}.
More recent controllers extend these methods to uncertain and dynamic environments.
Deep reinforcement learning trains a single policy that keeps a formation and avoids collisions at the same time~\cite{Sui2021-zl}.
Model-predictive control predicts the trajectories of moving obstacles and plans avoidance over a short horizon~\cite{Lindqvist2020-np}.
The same approach carries over to tasks beyond navigation, such as forming artistic patterns~\cite{5980269} and pushing objects collectively~\cite{10080994}.

Other work prepares a single swarm for many tasks, either by tuning its motion parameters or by training policies that switch among behaviors.
One approach adapted the attraction, repulsion, and preferred spacing of force-based motion rules to each task through automated search~\cite{engebraaten2020multifunction}.
Learning-based methods, on the other hand, train a single policy that produces several behaviors.
Through reinforcement learning, a swarm acquired a policy that can perform multiple tasks such as flocking, formation keeping, obstacle avoidance, and pursuit-evasion~\cite{batra2022decentralized}.
Imitation of an expert likewise gave real drones a single policy for flocking, leader following, and obstacle avoidance~\cite{chen2024stgnn}.

Fluidic Body brings this idea into human-led operation, where a person adjusts the parameters during the task.
This requires parameters that a person can change intuitively, unlike the gains of force-based rules or the weights of a learned policy.
Fluidic Body meets this requirement by using the viscosity and density of a fluid as its parameters, properties that people know from everyday materials.
We review work that builds swarm control on such physical models, and on fluids in particular, in \cref{subsec:physics_fluid}.

\subsection{Swarm Operation Methods}
\label{subsec:hsi}
Whatever controller drives the robots, a person can operate the swarm through many different methods depending on the application.
Research on human-swarm interaction has developed such methods for directing a swarm as a group, because steering each robot becomes too demanding as the swarm grows~\cite{kolling2016survey, kaduk2024onetomany}.
Many methods issue commands through a console, a master device, or natural language, for example by placing beacons~\cite{kolling2013foraging}, assigning tasks to groups~\cite{kolling2012control}, teleoperating the group~\cite{franchi2012bilateral}, or describing a target pattern~\cite{liu2024lgpf}.
Gestures in particular support diverse and intuitive input~\cite{nagi2014spatial, serpiva2021swarmpaint, giusti2012gesture, podevijn2014subswarms, suresh2020interpreter, kim2020userdefined}, and one method even lets a person treat the swarm as part of their own body~\cite{10.1145/3613904.3642870}.
Across these methods, the operator mostly specifies geometry, such as a place for the swarm to reach or a shape for it to hold.
Geometry alone, however, cannot adapt the swarm to tasks that demand different behaviors, such as flowing around obstacles and pushing objects.

Fluidic Body offers an operation method in which a person adjusts the swarm's behavior, not only its destination, by changing the fluid's properties through hand movements.
Because people understand how materials move and deform, interfaces grounded in such everyday physics are quick to learn~\cite{battaglia2013physics, jacob2008rbi}.

\subsection{Physics-inspired and Fluid-based Control}
\label{subsec:physics_fluid}
A line of research controls a swarm by modeling the collective itself as a physical medium, so that simple local rules give rise to the global behavior a designer wants~\cite{Spears2004-kw}.
Fluids suit this approach, since varying a few physical properties such as viscosity and density produces a wide range of collective behaviors.
Hydrodynamic formulations, and smoothed particle hydrodynamics (SPH)~\cite{Gingold1977-po} in particular, have been carried from fluid simulation to swarm control~\cite{Pimenta2008-ls, Pac2007-op}.
When each robot is treated as a fluid particle, the forces it exchanges with nearby robots make the swarm move as a fluid.
On this basis, the fluid model has been used to regulate how densely a swarm packs and to coordinate distributed agents~\cite{Zhao2011-nn, Shaw2011-wy}.
In more applied directions, fluid controllers led a swarm into formations that imitate human hand preshapes~\cite{Tilki2011-ky} and moved a swarm past obstacles without an explicit map of them~\cite{Eguchi18062025}.

Fluidic Body ties changes in the fluid's properties to human operation so that the control adapts to the task at hand.
Building on that controller~\cite{Eguchi18062025}, it exposes viscosity and density as interactive controls, so that a person switches between formations with different properties in real time through hand movements.

\section{Method}

\subsection{Overview}
\begin{figure*}[!t]
  \centering
  \includegraphics[width=1.0\linewidth]{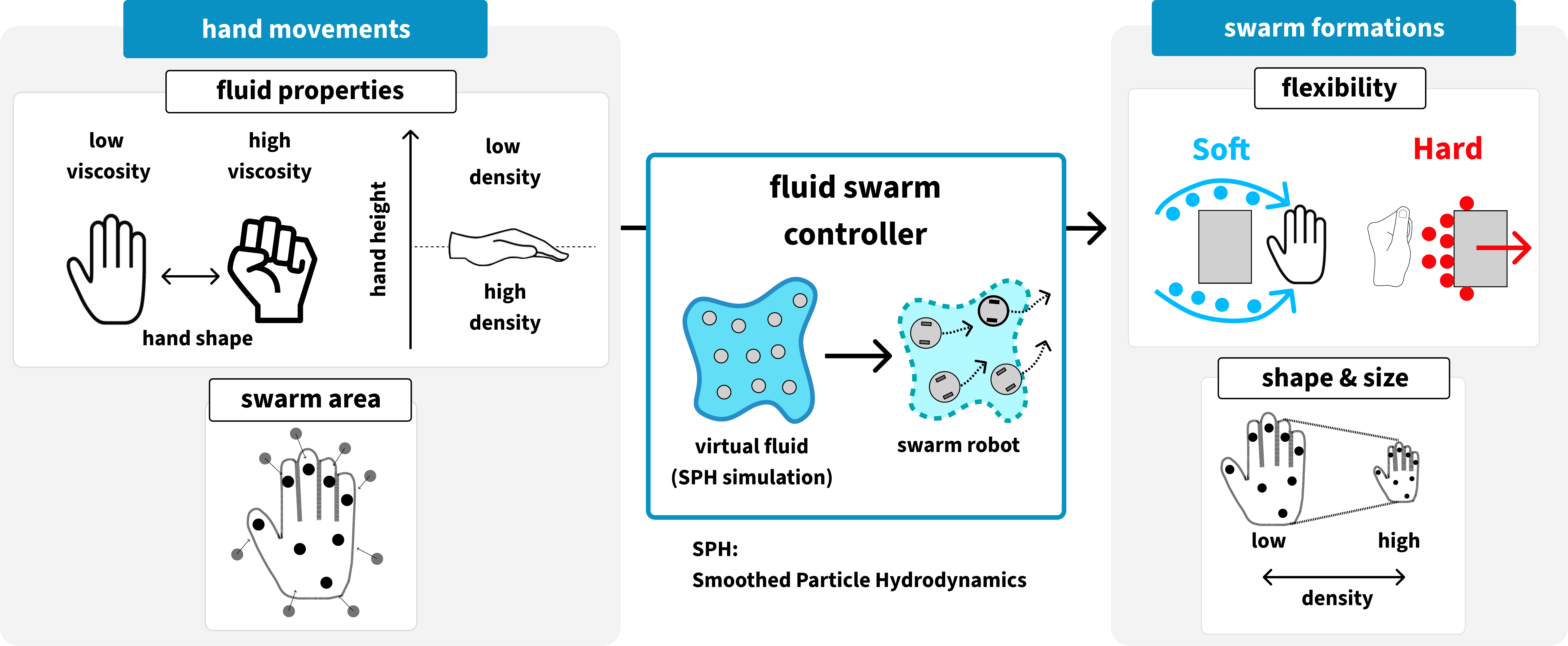}
  \caption{System design of Fluidic Body. The controller takes fluid properties and the swarm area as inputs. These inputs are determined by the user's hand shape, height, and outline, which are captured by a hand-tracking camera. The controller simulates a virtual fluid with SPH and treats each robot as one of its particles. Changing these inputs therefore changes the flexibility, shape, and size of the swarm.}
  \Description{System design of Fluidic Body. The controller takes fluid properties and the swarm area as inputs. These inputs are determined by the user's hand shape, height, and outline, which are captured by a hand-tracking camera. The controller simulates a virtual fluid with SPH and treats each robot as one of its particles. Changing these inputs therefore changes the flexibility, shape, and size of the swarm.}
  \label{fig:system_config}
\end{figure*}

Fluidic Body changes the behavior of a robot swarm according to the user's hand movements (\cref{fig:system_config}).
Its controller uses SPH~\cite{Gingold1977-po}, a method common in fluid simulators, to make the swarm imitate fluid motion~\cite{Pac2007-op}.
The controller takes the fluid properties, set by hand movements, and the swarm area as inputs.
It then outputs a velocity command to each robot, which changes the flexibility, shape, and size of the swarm.
We built its SPH formulation on that of our previous controller~\cite{Eguchi18062025, 10.1145/3721257.3734028} and modified parts of it.

This section describes the controller and the mapping from the hand to it separately.
\cref{subsec:controller} covers the controller itself, from the SPH forces that hold the swarm together to the shape force that fills the swarm area, and shows which parameters govern the flexibility and the spread of the formation.
\cref{subsec:mapping} then covers how hand movements set those parameters and the swarm area.

\subsection{Fluid Swarm Controller}
\label{subsec:controller}
The fluid swarm controller treats each robot $i$ as a particle of the fluid that SPH simulates.
In SPH, the interactions between particles decide where each particle moves.
Besides the fluid force given by SPH, the controller applies two more forces.
The velocity command $\bm{v}_i$ sent to each robot $i$ follows from the sum of the three, which we integrate with the forward Euler method at a sampling interval $\Delta t$:
\begin{equation} \label{eq:total_force}
    \frac{d\bm{v}_i}{dt} = \bm{f}^{\mathrm{sph}}_i + \bm{f}^{\mathrm{rep}}_i + \bm{f}^{\mathrm{sha}}_i,
\end{equation}
\begin{equation} \label{eq:euler}
    \bm{v}_i(t + \Delta t) = \bm{v}_i(t) + \frac{d\bm{v}_i(t)}{dt}\, \Delta t,
\end{equation}
where the norm of $\bm{v}_i$ is capped at the maximum robot speed $V_{\max}$.
The fluid force $\bm{f}^{\mathrm{sph}}_i$ makes the swarm behave as a fluid, while the repulsion $\bm{f}^{\mathrm{rep}}_i$ prevents collisions between robots.
The shape force $\bm{f}^{\mathrm{sha}}_i$ draws each robot into the area the user specifies while steering it clear of obstacles.
By supplying only the fluid properties and the swarm area, the user obtains a swarm that fills that area and behaves as the specified fluid would.
We describe the three forces in turn.

\subsubsection{Fluid Force}
SPH represents a fluid as a set of particles, each carrying the physical quantities of the fluid at its own position.
When the controller computes the properties of the fluid at a robot, it therefore takes a weighted sum over the robots around it.
The weights come from a smoothing kernel $W(R_{ij})$, for which we use a Gaussian over the robots within a range $\kappa h$:
\begin{equation} \label{eq:sph_kernel}
    W(R_{ij}) =
    \begin{cases}
        \alpha_d\, e^{-R_{ij}^2} & \text{if } R_{ij} \leq \kappa \\
        0 & \text{otherwise,}
    \end{cases}
    \quad
    \alpha_d = \frac{1}{\pi h^2},
    \quad
    R_{ij} = \frac{\|\bm{q}_i - \bm{q}_j\|}{h},
\end{equation}
where $\bm{q}_i$ is the position of robot $i$, $h$ is the smoothing length, and we set $\kappa = 2$ as in the base controller.
Each robot's density $\rho_i$ then follows from the masses $m_j$ of the neighboring robots:
\begin{equation} \label{eq:sph_density}
    \rho_i = \sum_{j=0}^{N-1} m_j\, W(R_{ij}),
\end{equation}
where $N$ is the number of robots.

The fluid force on robot $i$ combines a pressure term and a viscosity term:
\begin{equation} \label{eq:sph_force}
    \bm{f}^{\mathrm{sph}}_i = -\sum_{j=0}^{N-1} m_j \left( \frac{p_i}{\rho_i^2} + \frac{p_j}{\rho_j^2} \right) \nabla_i W(R_{ij}) + \bm{f}^{\mathrm{vis}}_i,
\end{equation}
where $\nabla_i W(R_{ij})$ is the gradient of the kernel, $p_i$ is the fluid pressure, and $\bm{f}^{\mathrm{vis}}_i$ is the viscosity term defined below.
Because we model an incompressible fluid, the pressure grows with the local density relative to a reference density $\rho_0$:
\begin{equation} \label{eq:sph_pressure}
    p_i = K \rho_0 \left[ \left( \frac{\rho_i}{\rho_0} \right)^{\gamma} - 1 \right],
\end{equation}
with stiffness constant $K$, reference density $\rho_0$, and adiabatic constant $\gamma = 7$ following~\cite{Zhao2011-nn}.

The viscosity term aligns the velocities across the swarm through the viscosity coefficient $\mu$, following~\cite{10.5555/846276.846298}:
\begin{equation} \label{eq:sph_viscosity}
    \bm{f}^{\mathrm{vis}}_i = \mu \sum_{j=0}^{N-1} m_j\, \frac{\bm{v}_j - \bm{v}_i}{\rho_j} \nabla_i^2 W(R_{ij}),
\end{equation}
where $\bm{v}_i$ is the velocity of robot $i$ and $\nabla_i^2 W(R_{ij})$ is the Laplacian of the kernel.

Among these parameters, the user mainly controls two.
A larger viscosity coefficient $\mu$ in \cref{eq:sph_viscosity} aligns the robot velocities more strongly.
The formation then resists deformation and moves as a stiff, cohesive mass.
With a smaller $\mu$, the formation deforms and slips through narrow gaps.
The reference density $\rho_0$ in \cref{eq:sph_pressure} sets the target spacing between robots and therefore governs how compact or spread out the formation is.
By adjusting $\mu$ and $\rho_0$ during operation, the user softens the formation for avoidance and stiffens it for transport.

\subsubsection{Inter-robot Repulsion}
The repulsion keeps neighboring robots from running into each other:
\begin{equation} \label{eq:rep_force}
    \bm{f}^{\mathrm{rep}}_i = K_{\mathrm{rep}} \sum_{j=0}^{N-1} \frac{1}{(\|\bm{q}_i - \bm{q}_j\| - 2r)^2} \nabla_i W(R_{ij}),
\end{equation}
where $K_{\mathrm{rep}}$ is the repulsion gain and $r$ is the radius of a robot.

\subsubsection{Shape Force}
\label{subsubsec:shape_force}
\begin{figure}[t]
  \centering
  \includegraphics[width=\linewidth]{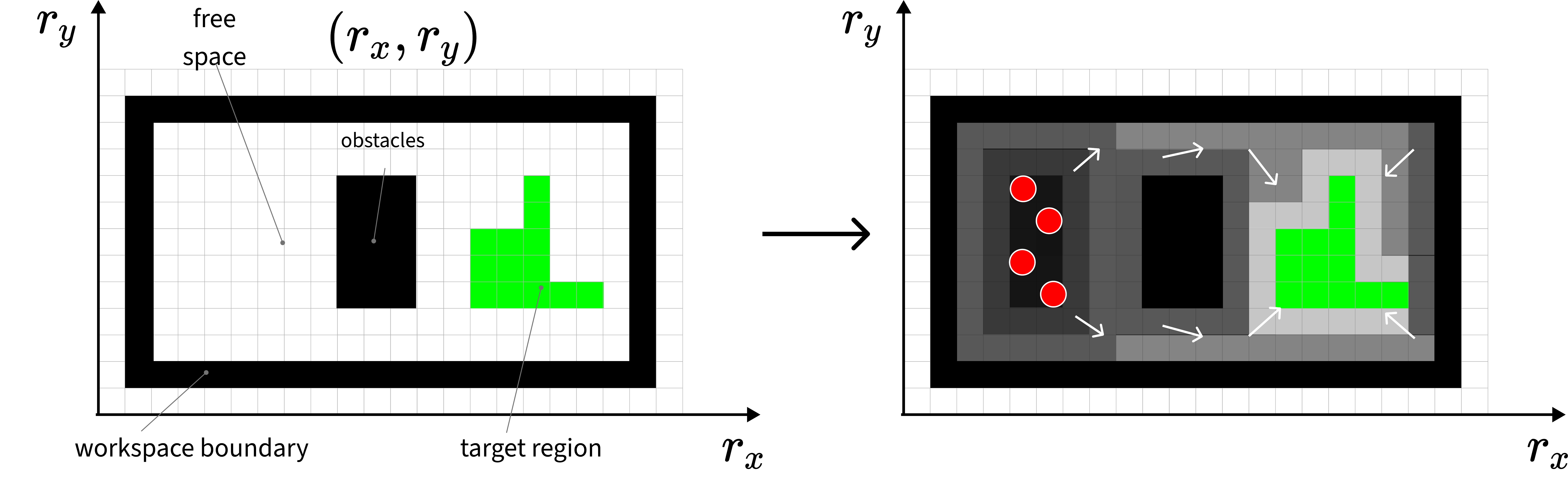}
  \caption{How the shape force is obtained. Left: the swarm area and the obstacles are written on a common grid. The grid splits the workspace into the target region $\mathcal{S}$ in green, the obstacles $\mathcal{O}$ and the workspace boundary $\partial \mathcal{W}$ in black, and the free space $\mathcal{F}$ in white. Right: the potential field solved on that grid, light where it is low next to the target region and dark where it is high next to the obstacles. Each robot, drawn in red, follows the steepest descent of the field, and the arrows show how this carries the swarm around the obstacle and into the target region.}
  \Description{Two panels. The left panel shows a rectangular workspace on a square grid, with a black border for the workspace boundary, a black rectangle for an obstacle in the middle, a green cluster of cells on the right for the target region, and white cells elsewhere for the free space. The right panel shows the same layout filled with a grayscale potential field, light near the green target region and dark away from it, with white arrows pointing from the robots on the left, around the obstacle, toward the target region.}
  \label{fig:shape_force}
\end{figure}

The swarm area, which the user marks out with the outline of their hand, defines the formation shape of the swarm, and the shape force draws the robots into it.
The other two forces are computed by each robot from the robots near it, whereas this one is planned globally over the whole workspace, obstacles included.
We write this area and the obstacles on a common grid, where the cells inside the area form the target region $\mathcal{S}$, the cells a physical obstacle occupies form the obstacle set $\mathcal{O}$, and the rest form the free space $\mathcal{F}$.
We lay a potential field $u$ over the free space, low at the target region and high at the obstacles, and let each robot follow it downhill (\cref{fig:shape_force}).
The values in between come from Laplace's equation, which fills the free space as smoothly as those two boundaries allow:
\begin{equation} \label{eq:laplace}
    \begin{cases}
        \nabla^2 u(\bm{r}) = 0, & \bm{r} \in \mathcal{F} \quad \text{(free space)},\\[2pt]
        u(\bm{r}) = 0,          & \bm{r} \in \mathcal{S} \quad \text{(target region)},\\[2pt]
        u(\bm{r}) = 1,          & \bm{r} \in \mathcal{O} \quad \text{(obstacles)},\\[2pt]
        \partial u / \partial n = 0, & \bm{r} \in \partial \mathcal{W} \quad \text{(workspace boundary)},
    \end{cases}
\end{equation}
where $\bm{r} = (r_x, r_y)$ is a grid position, and the last condition keeps the outer boundary $\partial \mathcal{W}$ of the workspace from pushing the robots the way an obstacle does.
Solving \cref{eq:laplace} amounts to replacing the value of every free cell with the average of its four neighbors until the values settle.
We repeat this once per control cycle so that the field follows the swarm area as the user moves their hand.
A field of this kind takes its extreme values only on the boundaries.
The free space therefore holds no dead end where a robot could settle short of the target region~\cite{Connolly1990-laplace, Kim1992-harmonic}.

The shape force follows the steepest descent of this field.
Since the field decays sharply away from the target region, we take only the direction of that descent and drop the force once a robot is inside the target region:
\begin{equation} \label{eq:shape_force}
    \bm{f}^{\mathrm{sha}}_i =
    \begin{cases}
        \bm{0} & \text{if } \bm{q}_i \in \mathcal{S} \\[4pt]
        -K_{\mathrm{sha}} \dfrac{\nabla u(\bm{q}_i)}{\|\nabla u(\bm{q}_i)\|} & \text{otherwise,}
    \end{cases}
\end{equation}
where $\bm{q}_i$ is the position of robot $i$ and $K_{\mathrm{sha}}$ is the shape gain.
Zeroing the force inside the region turns path following into shape filling.
The field only has to deliver each robot into $\mathcal{S}$ past the obstacles.
From there, the fluid force of \cref{eq:sph_force} takes over.
It spreads the robots at the spacing set by $\rho_0$ until they fill the region evenly, and it holds that arrangement as rigidly as $\mu$ dictates.
The same mechanism transports the swarm, because moving the swarm area moves the sink of the field and the swarm flows after it around whatever obstacles lie on the way.

\subsection{Hand Operation Mapping}
\label{subsec:mapping}
We map the movements of the hand onto the inputs of the fluid swarm controller, so that a person handles the swarm intuitively with their hand.
A hand-tracking camera gives the shape, the height, and the outline of the hand, and each of the three drives one quantity the controller takes.

The shape of the hand switches the fluid between two settings, hard and soft, which differ both in the viscosity coefficient $\mu$ of \cref{eq:sph_viscosity} and in the reference density $\rho_0$ of \cref{eq:sph_pressure}.
An open hand selects the soft setting.
Its low $\mu$ and low $\rho_0$ let the formation deform and spread so that it slips through narrow gaps.
A closed hand selects the hard setting, whose high $\mu$ and high $\rho_0$ hold the robots together as a compact mass that pushes objects.
The height of the hand above the table also acts on $\rho_0$: raising the hand lowers it, which widens the spacing between the robots and spreads the formation.
The outline of the hand becomes the swarm area, the target region $\mathcal{S}$ of \cref{eq:shape_force}, and the formation follows the outline as it changes.

Because a person can combine the three movements, they can adapt the behavior of the swarm to the task while operating it.

\section{Technical Evaluation}
We first evaluated the proposed controller in simulation, with automated control in place of a human operator.
We compared the proposed and baseline methods on success rate and completion time across several environments with different characteristics.

This evaluation asks:

\noindent{\itshape\textbf{Q1.} Does matching the fluid properties to each task let the proposed method adapt better than a fixed baseline that keeps the same behavior for every task?}

\begin{figure}[t]
  \centering
  \includegraphics[width=\linewidth]{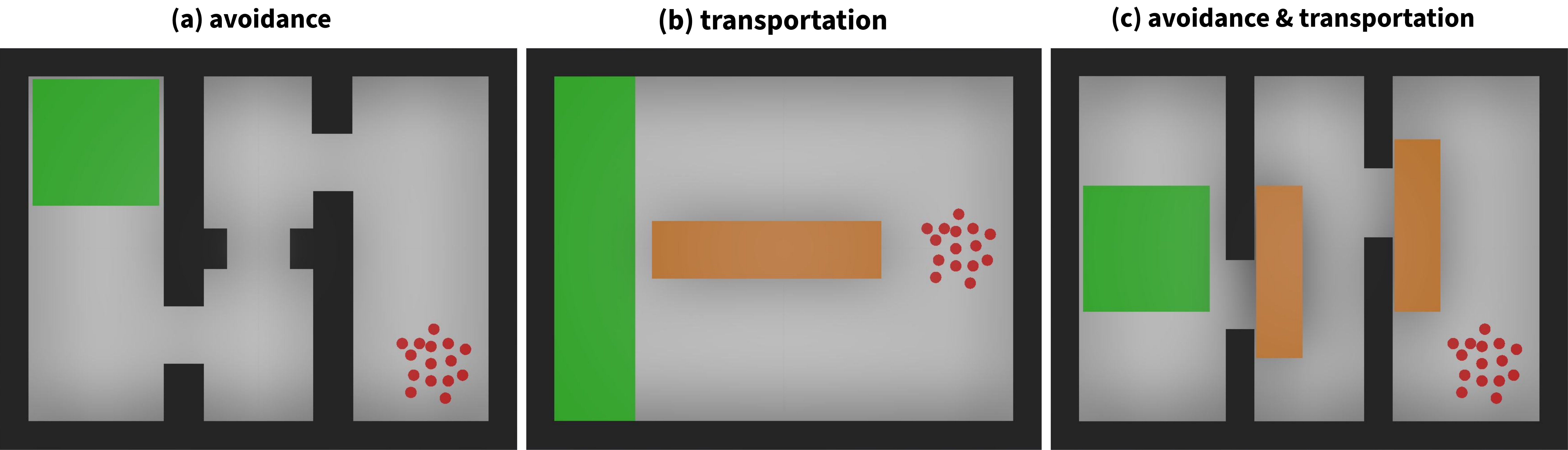}
  \caption{The three simulation environments: (a) obstacle avoidance, (b) object transportation, and (c) avoidance and transportation. The technical evaluation uses (a) and (b), and the user study uses all three. The swarm robots are red, the static obstacles are black, the movable object is orange, and the green frame is the goal. In obstacle avoidance, the swarm passes the obstacles and gathers all robots inside the goal frame ($0.2 \times 0.2$~m). In object transportation, the swarm moves the object into the goal frame ($0.2 \times 0.6$~m). In the avoidance and transportation environment, the swarm moves an object out of the way to open a path through the obstacles and gathers all robots inside the goal frame ($0.2 \times 0.2$~m) within a single trial.}
  \Description{The three simulation environments: (a) obstacle avoidance, (b) object transportation, and (c) avoidance and transportation. The technical evaluation uses (a) and (b), and the user study uses all three. The swarm robots are red, the static obstacles are black, the movable object is orange, and the green frame is the goal. In obstacle avoidance, the swarm passes the obstacles and gathers all robots inside the goal frame. In object transportation, the swarm moves the object into the goal frame. In the avoidance and transportation environment, the swarm moves an object out of the way to open a path through the obstacles and gathers all robots inside the goal frame within a single trial.}
  \label{fig:tech_eval_env}
\end{figure}

\subsection{Tasks and Simulation Environment}
To test the controller on tasks with contrasting requirements, we designed two tasks following the task design of prior multi-robot interaction research~\cite{10161412} (\cref{fig:tech_eval_env}):

\begin{itemize}[leftmargin=4mm]
\item \textbf{Obstacle Avoidance} (\cref{fig:tech_eval_env}a): the swarm must pass obstacles and gather all robots inside a goal frame. A trial is successful when every robot is contained within the frame.
\item \textbf{Object Transportation} (\cref{fig:tech_eval_env}b): the swarm must move the orange object into a goal frame. A trial is successful when the object is fully placed inside the frame.
\end{itemize}
Both tasks share a $0.8 \times 0.6$~m field, and each trial has a time limit of 60~s.

We ran these tasks in a simulator built on the Unity game engine.
The simulator ran on a laptop PC with 36~GB of RAM, an Intel Core i7-12650H CPU, and a GeForce RTX 4060 Laptop GPU with 8~GB of VRAM.
The swarm consists of 16 homogeneous robots, each 30~mm in diameter with a maximum travel speed of 40~mm/s.
Each robot is omnidirectional and moves along the two-dimensional velocity vector that the controller outputs.
The simulator provides the ground-truth position of every robot and obstacle to the controller at each time step.

\subsection{Conditions}
We compared three controller conditions:

\begin{itemize}[leftmargin=4mm]
\item \textbf{RVO}: the reciprocal velocity obstacle method~\cite{Van_den_Berg2008-we}. It computes a collision-free velocity command for each robot from the velocities of the surrounding robots and obstacles. Each robot thus predicts and avoids collisions while heading to the goal. Many swarm robot interfaces build on RVO~\cite{Le_Goc2016-dg, 10.1145/3613904.3642870}, because its simple mechanism already covers both inter-robot avoidance and obstacles of simple shape. Unlike our controller, RVO uses the same fixed parameters for every task.
\item \textbf{Fluid soft (Ours)}: the proposed fluid-imitation controller tuned to a soft, easily deforming formation.
\item \textbf{Fluid hard (Ours)}: the proposed controller tuned to a stiff, cohesive formation.
\end{itemize}
We set the soft and hard conditions by adjusting several fluid parameters together, including the viscosity and density, so that the formation either deforms freely or holds a cohesive shape. Each condition used a single parameter setting that we fixed in advance, and no trial switched between soft and hard. We chose both settings in pilot trials and kept the values that performed best. \Cref{tab:params} lists the fluid parameter values for the two conditions, and we used the standard RVO parameters~\cite{Van_den_Berg2008-we} for the baseline.

\begin{table}[t]
  \centering
  \caption{Fluid parameter values for the soft and hard conditions in the technical evaluation.}
  \label{tab:params}
  \begin{tabular}{lcc}
    \toprule
    Parameter & Fluid soft & Fluid hard \\
    \midrule
    Viscosity coefficient $\mu$      & 0.001 & 0.03 \\
    Reference density $\rho_0$       & 0.4 & 0.6 \\
    \bottomrule
  \end{tabular}
\end{table}

\subsection{Procedure and Metrics}
We ran 50 trials for each condition on each task.
At the start of every trial, each robot was placed randomly around its initially configured position.

Because this evaluation had no human operator, we controlled the swarm with a sequence of waypoints that we scripted in advance. These waypoints stood in for the goal position that a hand would otherwise give. The waypoints moved at a constant 0.1~m/s and were identical across all conditions, and at each step we passed the current waypoint to the swarm as its goal position. This gave every controller the same global path for a task and left only the local behavior to the controller. Any difference in success therefore came from how each controller adapted its formation rather than from the route it took.

We measured two metrics: the \emph{success rate}, defined as the proportion of trials completed within the 60~s limit, and the \emph{completion time}, defined as the time to satisfy the success condition. The completion times we report as a mean $\pm$ SD cover the completed trials only.

\begin{table}[t]
  \centering
  \caption{Success rate and completion time for each control condition on the obstacle avoidance and object transportation tasks. Success rate is given as a percentage with the number of successful trials in parentheses, and completion time is the mean $\pm$ SD over the completed trials only. \Cref{fig:tech_eval_results} reports the accompanying statistical tests.}
  \label{tab:tech_eval}
  \begin{tabular}{lcccc}
    \toprule
    & \multicolumn{2}{c}{Avoidance} & \multicolumn{2}{c}{Transportation} \\
    \cmidrule(lr){2-3} \cmidrule(lr){4-5}
    Method & Success (\%) & Time (s) & Success (\%) & Time (s) \\
    \midrule
    RVO                  & 16.0 (8/50)   & $20.41 \pm 0.30$ & \textbf{100.0 (50/50)} & $37.77 \pm 2.01$ \\
    Fluid soft (Ours) & \textbf{98.0 (49/50)}  & $20.56 \pm 0.60$ & 36.0 (18/50)  & $45.11 \pm 3.05$ \\
    Fluid hard (Ours) & 70.0 (35/50)  & $19.76 \pm 0.13$ & \textbf{100.0 (50/50)}  & $36.05 \pm 0.50$ \\
    \bottomrule
  \end{tabular}
\end{table}

\begin{figure*}[t]
  \centering
  \includegraphics[width=\textwidth]{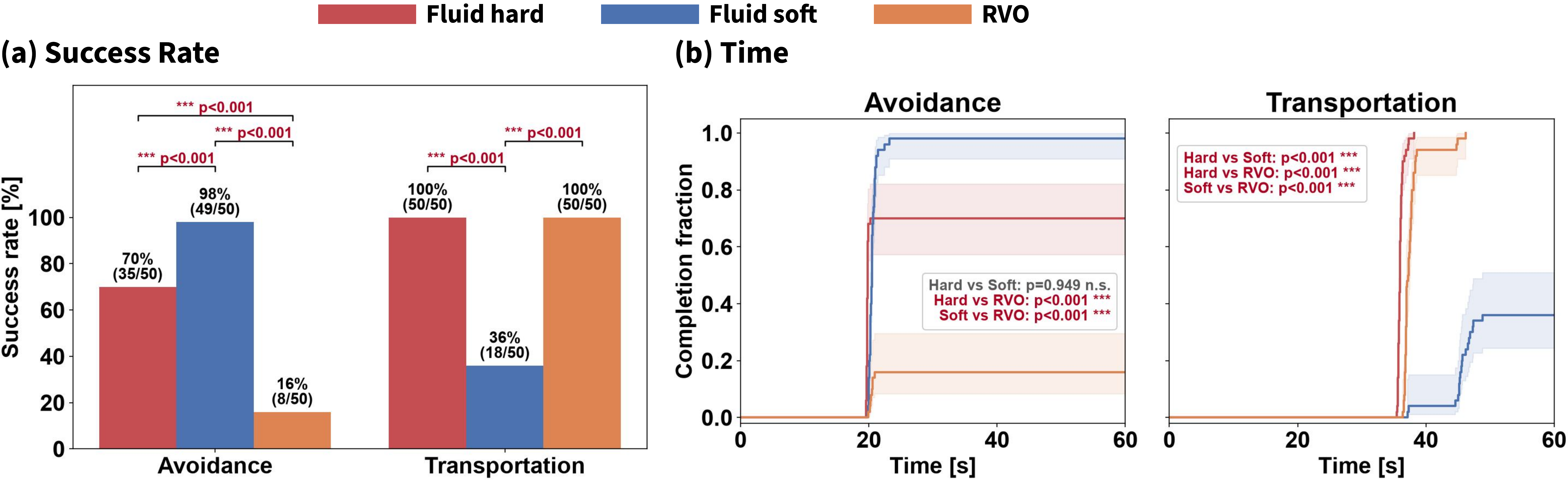}
  \caption{Success rate and completion time for each controller on the two tasks: (a) success rate, and (b) Kaplan-Meier completion curves, where a timed-out trial is right-censored and the shaded band is the 95\% confidence interval. Significance markers: $^{***}p<.001$; n.s.\ denotes no significant difference.}
  \Description{Two panels comparing Fluid hard, Fluid soft, and RVO on the avoidance and transportation tasks: (a) success rate and (b) Kaplan-Meier completion curves against time. Fluid soft completes the most avoidance trials, while RVO and Fluid hard complete every transportation trial and Fluid hard completes them fastest.}
  \label{fig:tech_eval_results}
\end{figure*}

\subsection{Results}
\Cref{tab:tech_eval} summarizes the results, and \cref{fig:tech_eval_results} shows the success rates and the Kaplan-Meier completion curves together with the corresponding tests.
Trials ran independently and were not paired across controllers.
We therefore compared the three controllers pairwise within each task as independent groups.
Below, we report the results one metric at a time and give the analysis for each metric alongside its results.
Every $p$ value below is Holm-adjusted over the six comparisons that two tasks and three controller pairs produce.

\subsubsection{Success rate}
Because the trials of different controllers were independent, we compared every controller pair within a task with a two-sided Fisher's exact test and report Cohen's $h$ as the effect size.
On the avoidance task, Fluid soft completed the most trials (49 of 50, 98\%), followed by Fluid hard (35 of 50, 70\%) and RVO (8 of 50, 16\%).
Every pair differed significantly ($p < .001$ for all three; $h = 0.88$ for Fluid soft vs.\ Fluid hard, $h = 1.16$ for Fluid hard vs.\ RVO, $h = 2.03$ for Fluid soft vs.\ RVO).
On transportation, RVO and Fluid hard both completed all 50 trials (100\%, $p = 1.000$, $h = 0$ between them).
Fluid soft completed only 18 of 50 (36\%) and fell significantly below both ($p < .001$, $h = 1.85$ against each).

\subsubsection{Completion time}
Analyzing only the trials that finished would discard the timeouts. We therefore analyzed all 50 trials with survival analysis. Task completion was the event, and a trial that reached the 60~s limit was right-censored. We compared the Kaplan-Meier completion curves pairwise with a log-rank test. As the effect size, we report the hazard ratio (HR) of a Cox proportional-hazards model with its 95\% confidence interval (CI). An HR above 1 indicates that the first controller of a pair completed the task earlier or more often than the second.

Completion times over the completed trials were nearly identical across conditions on the avoidance task ($20.41 \pm 0.30$~s for RVO, $20.56 \pm 0.60$~s for Fluid soft, and $19.76 \pm 0.13$~s for Fluid hard). Note that on this task the swarm followed a waypoint that advanced at a constant 0.1~m/s, and the completion time therefore reflected the waypoint pace more than the controller.
The survival analysis on this task thus separates the controllers by their completion rate rather than their completion time. Both fluid settings completed significantly more of their trials than RVO (log-rank $p < .001$ for each; Fluid hard, HR $= 8.20$, 95\% CI $[3.77, 17.83]$; Fluid soft, HR $= 15.76$, 95\% CI $[7.20, 34.50]$).
The two fluid settings did not differ (log-rank $p = .949$).

On transportation, the controllers differed in completion time. Fluid hard finished fastest ($36.05 \pm 0.50$~s), ahead of RVO ($37.77 \pm 2.01$~s), and the survival analysis confirms this gap (log-rank $p < .001$; HR $= 6.06$, 95\% CI $[3.87, 9.51]$).
Fluid soft fell significantly behind both (log-rank $p < .001$ against each; Fluid hard vs.\ Fluid soft, HR $= 115.38$, 95\% CI $[26.54, 501.71]$; Fluid soft vs.\ RVO, HR $= 0.04$, 95\% CI $[0.02, 0.08]$).

Overall, selecting the fluid setting suited to each task gave the best result on both tasks. Fluid soft was the most reliable on avoidance, while Fluid hard matched RVO in completing every transportation trial and finished significantly faster. The fixed baseline, by contrast, was reliable on only one of the two tasks. These results answer Q1 positively, and we discuss the task-dependent trade-off behind them in \cref{subsec:disc_q1}.

\section{User Study}
In this study, participants operated the swarm themselves. We compared our controller against a baseline on task performance and the operators' subjective experience across several environments.

This study asks:

\noindent{\itshape\textbf{Q2.} Can users complete a variety of tasks faster and more reliably than a baseline by changing how the swarm moves through fluid properties?}

\smallskip
\noindent{\itshape\textbf{Q3.} Does changing how the swarm moves through fluid properties improve the user's operating experience over the baseline, and does this experience contribute to the efficiency of operation?}

\subsection{Participants}
We recruited 17 right-handed participants online (13 men and 4 women, mean age $25.3 \pm 11.4$ years).
Thirteen of them had prior experience with VR.
Each session lasted about one hour, and every participant received monetary compensation for their time.
The study was approved by the Cluster, Inc. Research Ethics Committee (registration number: 2025-003-01).

\subsection{Conditions and Tasks}
We used a within-participant design that crossed two controllers with three environments (\cref{fig:tech_eval_env}).
Crossing the two controllers with the three environments produced six tasks per participant.

\subsubsection{Controllers}
\begin{itemize}[leftmargin=4mm]
\item \textbf{RVO}: following Swarm Body~\cite{10.1145/3613904.3642870}, one robot is assigned to each joint of the hand. The swarm thus follows the shape and motion of the hand directly, while RVO keeps the robots collision-free.
\item \textbf{Fluid (Ours)}: the robots spread evenly to fill the swarm area given by the outline of the hand, and the operator switches between the soft and hard settings from the technical evaluation by how far the hand is opened. The robots turn blue for the soft setting and red for the hard setting to show the current state.
\end{itemize}
Both conditions used 16 robots, as in the technical evaluation.

\subsubsection{Environments}
\begin{itemize}[leftmargin=4mm]
\item \textbf{Obstacle Avoidance} (\cref{fig:tech_eval_env}a): the participant guides the swarm past the obstacles and gathers all robots inside the goal frame.
\item \textbf{Object Transportation} (\cref{fig:tech_eval_env}b): the participant moves the object into the goal frame.
\item \textbf{Avoidance and Transportation} (\cref{fig:tech_eval_env}c): the participant moves an object out of the way to open a path through the obstacles and gathers all robots inside the goal frame within a single trial.
\end{itemize}

\subsection{Apparatus}
\begin{figure}[t]
  \centering
  \includegraphics[width=\linewidth]{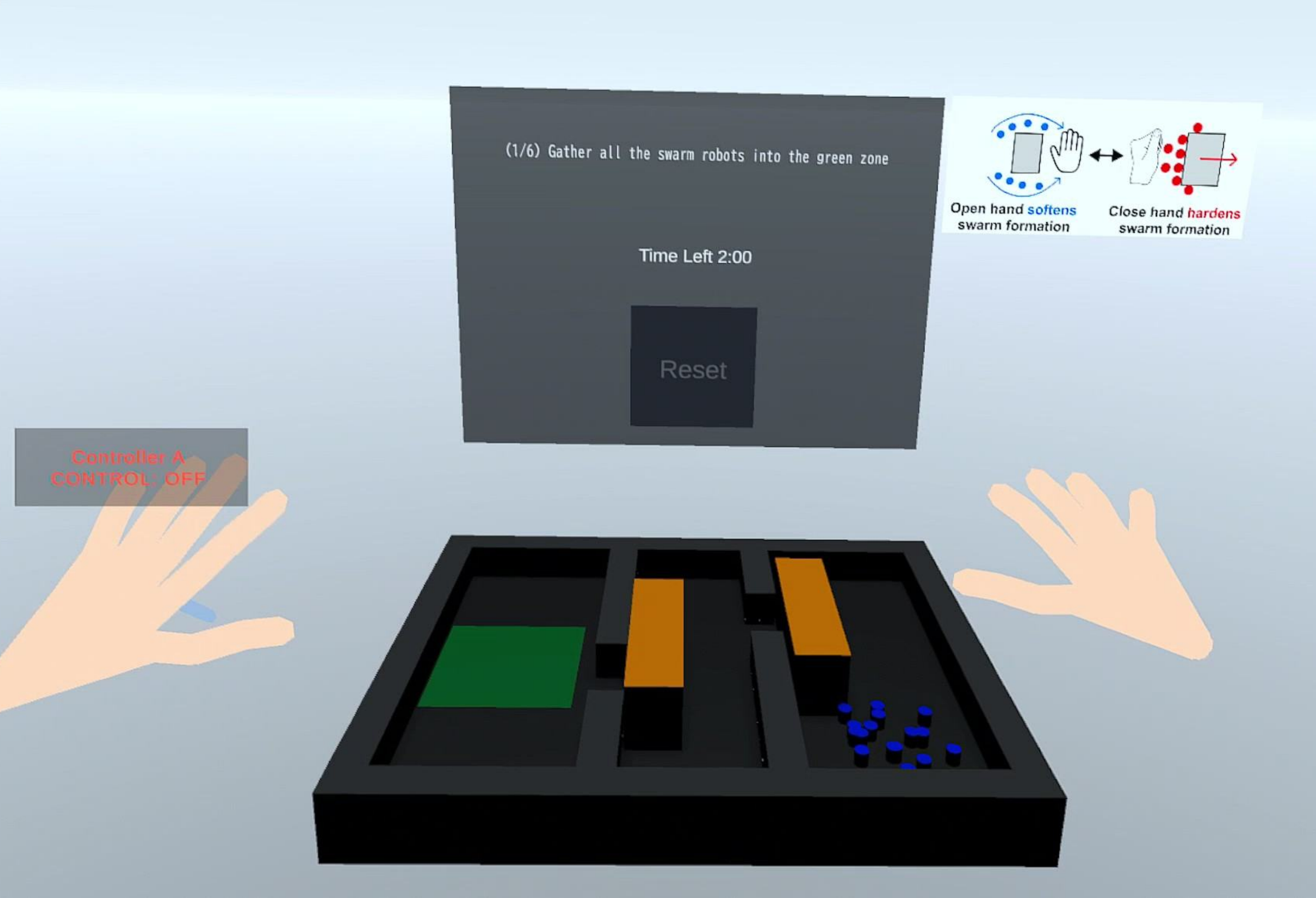}
  \caption{First-person view of the user study in VR. The participant guides the blue swarm robots with tracked hands toward the green goal zone while dealing with the orange obstacles and transport objects. The panel shows the current task and the remaining time, and the top-right reminder shows which controller is active.}
  \Description{A first-person VR screenshot showing two tracked hands, a boxed task field with a green goal zone, orange objects, and blue swarm robots, an instruction panel with the task text and a timer, and a top-right reminder showing the active controller.}
  \label{fig:user_study_vr}
\end{figure}

Participants operated the swarm inside a Unity environment, which they viewed through a Meta Quest~3 headset in wired PCVR mode (\cref{fig:user_study_vr}).
The same laptop PC as in the technical evaluation ran this environment.
Participants wore noise-canceling headphones that played pink noise to mask external sounds, and those who wore glasses kept them on inside the headset.

The participant saw a boxed field on the table with a green goal zone, orange obstacles and transport objects, and the blue swarm robots, together with their own tracked hands.
An instruction panel above the field showed the current task and the remaining time, and a reminder in the top-right corner showed which controller the participant was operating.

Participants operated the swarm with their right hand only and used their left hand for the interface, which is why we recruited right-handed participants.
Through this interface, they could press a reset button to return the robots and the obstacle layout to their initial positions while the timer kept running.

\subsection{Procedure}
At the start of the session, we explained the procedure and the tasks to the participant and obtained written informed consent.
We did not disclose the hypothesis of the study.
To keep participants from recognizing which controller was our proposed method, we presented the two controllers as Controller A and Controller B.
Which controller each label denoted varied across participants.
For each controller, participants first practiced for about 10 minutes in a separate training environment that held simple obstacles and a transport object and that differed from the six evaluation tasks.
They then performed the evaluation tasks, one environment at a time. Each task had a time limit of 120~s, and a trial that reached it counted as a timeout. We counterbalanced the order of the controllers and environments across participants with a Latin square.
After each task, participants answered questionnaires on embodiment, agency, and workload inside the VR space, and they took a short break of about one minute after every two tasks.
At the end of the session, we ran a semi-structured interview about how they had felt while operating the swarm.

\subsection{Measures}
Following the technical evaluation, we recorded completion time and success rate as objective measures.
For subjective measures, participants answered an embodiment questionnaire and the NASA Task Load Index (NASA-TLX)~\cite{nasa-tlx} after every task.
The embodiment questionnaire, adopted from Swarm Body~\cite{10.1145/3613904.3642870}, had two subscales, body ownership and agency, with four items each (\cref{tab:questionnaire}).
Participants rated every item on a 7-point Likert scale (1~=~strongly disagree, 7~=~strongly agree) inside the VR space.
Following prior work~\cite{10.1145/3613904.3642870}, we computed the body ownership and agency scores as the mean of their four corresponding items.
For NASA-TLX, participants rated the six subscales after every task and completed the pairwise weight comparisons once after finishing all tasks.
We computed the weighted workload by applying each participant's weights to their six subscale ratings.
In the semi-structured interview, we collected qualitative feedback on the operation experience.

\begin{table}[t]
  \centering
  \caption{Items of the embodiment questionnaire~\cite{10.1145/3613904.3642870}, grouped by subscale.}
  \label{tab:questionnaire}
  \begin{tabular}{@{}p{0.22\linewidth}p{0.7\linewidth}@{}}
    \toprule
    Subscale & Item \\
    \midrule
    Body ownership
      & It felt like the swarm robot was my body. \\
      & It felt like some of the robots were my fingers. \\
      & It felt like the swarm robots belonged to me. \\
      & The swarm robot felt like a human hand. \\
    \midrule
    Agency
      & The movements of the swarm robot felt like they were my movements. \\
      & I felt like I was controlling the movements of the swarm robot. \\
      & I felt like I was causing the movements of the swarm robot. \\
      & The movements of the swarm robot were in sync with my own movements. \\
    \bottomrule
  \end{tabular}
\end{table}

\subsection{Results}
\Cref{tab:user_eval} reports success rate and completion time, \cref{fig:user_subjective} shows the objective results and the subjective ratings, and \cref{fig:user_trajectory} shows how participants moved the swarm in each task.
We compared our controller against the RVO baseline within each task.
Below, we report the results one measure at a time and give the analysis for each measure alongside its results.
Every $p$ value below is Holm-adjusted within its own family of tests.

\subsubsection{Success rate and completion time}
For success rate, we compared the two controllers within each task with an exact McNemar test and report Cohen's $g$ as the effect size.
Both controllers handled the avoidance task and the transportation task almost perfectly.
They reached the goal in every avoidance trial and succeeded in 16 of 17 transportation trials (94.1\%, $p = 1.000$, $g = 0$).
On the avoidance and transportation task, however, the gap widened.
Our controller succeeded in all 17 trials, whereas the RVO baseline succeeded in only 8 (47.1\%) and timed out in the rest (\cref{fig:user_subjective}a, \cref{tab:user_eval}; $p = .008$, $g = 0.50$).

For completion time, our controller was on average at least as fast as the RVO baseline on every task.
The gap was largest on the avoidance and transportation task ($38.5 \pm 19.9$~s vs.\ $76.9 \pm 34.2$~s; \cref{tab:user_eval}).
We analyzed completion time with survival analysis.
Task completion was the event, and timed-out trials were right-censored.
We compared the Kaplan-Meier completion curves with a log-rank test and report the HR with its 95\% CI as in the technical evaluation (HR $> 1$ favors our controller).
Participants finished faster with our controller on the avoidance and transportation task, where it reached a Kaplan-Meier median of 35.2~s while the RVO curve never reached 50\% completion (\cref{fig:user_subjective}b; log-rank $p < .001$; HR $= 10.25$, 95\% CI $[3.18, 33.02]$).
On the avoidance task, the difference between the two controllers was marginally significant (Kaplan-Meier median 14.9~s vs.\ 22.2~s; log-rank $p = .064$; HR $= 2.14$, 95\% CI $[0.96, 4.78]$).
On the transportation task we found no difference between the two controllers (Kaplan-Meier median 34.4~s vs.\ 27.7~s; log-rank $p = .912$; HR $= 1.04$, 95\% CI $[0.50, 2.15]$).

Overall, our controller was at least as fast as the baseline on the simpler tasks and clearly outperformed it on the combined avoidance and transportation task.
This gives a positive answer to Q2, most clearly on the hardest task, and we discuss the task-dependent pattern behind it in \cref{subsec:disc_q2}.

\subsubsection{Workload (NASA-TLX)}
We confirmed with a Shapiro-Wilk test that the within-participant differences were normal.
We then compared the two controllers with a paired $t$-test and report Cohen's $d_z$ as the effect size.

Our controller significantly reduced the weighted workload on the avoidance and transportation task (\cref{fig:user_subjective}c; $M = 30.3$, $SD = 14.9$ vs.\ $M = 62.7$, $SD = 21.3$; $t(16) = -6.12$, $p < .001$, $d_z = -1.49$).
On the avoidance task, workload did not differ significantly between the two controllers ($M = 20.1$, $SD = 13.1$ vs.\ $M = 28.1$, $SD = 15.5$; $t(16) = -1.97$, $p = .396$, $d_z = -0.48$).
We found no difference on the transportation task ($M = 33.0$, $SD = 17.3$ vs.\ $M = 35.4$, $SD = 19.2$; $t(16) = -0.48$, $p = 1.000$, $d_z = -0.12$).

\subsubsection{Body ownership}
As with workload, we compared the controllers with a paired $t$-test, and no task showed a significant difference.
On the avoidance and transportation task, ownership did not differ significantly between the two controllers (\cref{fig:user_subjective}d; $M = 4.99$, $SD = 1.29$ vs.\ $M = 4.03$, $SD = 0.87$; $t(16) = 2.64$, $p = .125$, $d_z = 0.64$).
The two controllers did not differ on the avoidance task ($M = 5.15$, $SD = 1.10$ vs.\ $M = 5.13$, $SD = 1.02$; $t(16) = 0.05$, $p = 1.000$, $d_z = 0.01$) and on the transportation task ($M = 4.87$, $SD = 1.34$ vs.\ $M = 4.94$, $SD = 1.37$; $t(16) = -0.21$, $p = 1.000$, $d_z = -0.05$).

\subsubsection{Agency}
We again compared the controllers with a paired $t$-test.
Agency was significantly higher with our controller on the avoidance and transportation task (\cref{fig:user_subjective}e; $M = 5.96$, $SD = 0.80$ vs.\ $M = 3.97$, $SD = 0.90$; $t(16) = 6.80$, $p < .001$, $d_z = 1.65$).
We found no difference between the two controllers on the avoidance task ($M = 5.91$, $SD = 0.80$ vs.\ $M = 5.57$, $SD = 0.89$; $t(16) = 1.41$, $p = .894$, $d_z = 0.34$) or the transportation task ($M = 5.47$, $SD = 1.33$ vs.\ $M = 5.49$, $SD = 1.08$; $t(16) = -0.05$, $p = 1.000$, $d_z = -0.01$).

Overall, the subjective measures followed the same pattern as efficiency and separated the controllers only on the combined task.
On that task, our controller lowered workload and raised agency, and ownership moved the same way without reaching significance.
On the two simpler tasks, we found no difference between the two controllers on any measure.
This gives a positive answer to Q3, carried by workload and agency, and we examine how this operating experience relates to the efficiency gain in \cref{subsec:disc_q3}.

\subsubsection{Trajectories}
\begin{figure*}[tp]
  \centering
  \includegraphics[width=\linewidth]{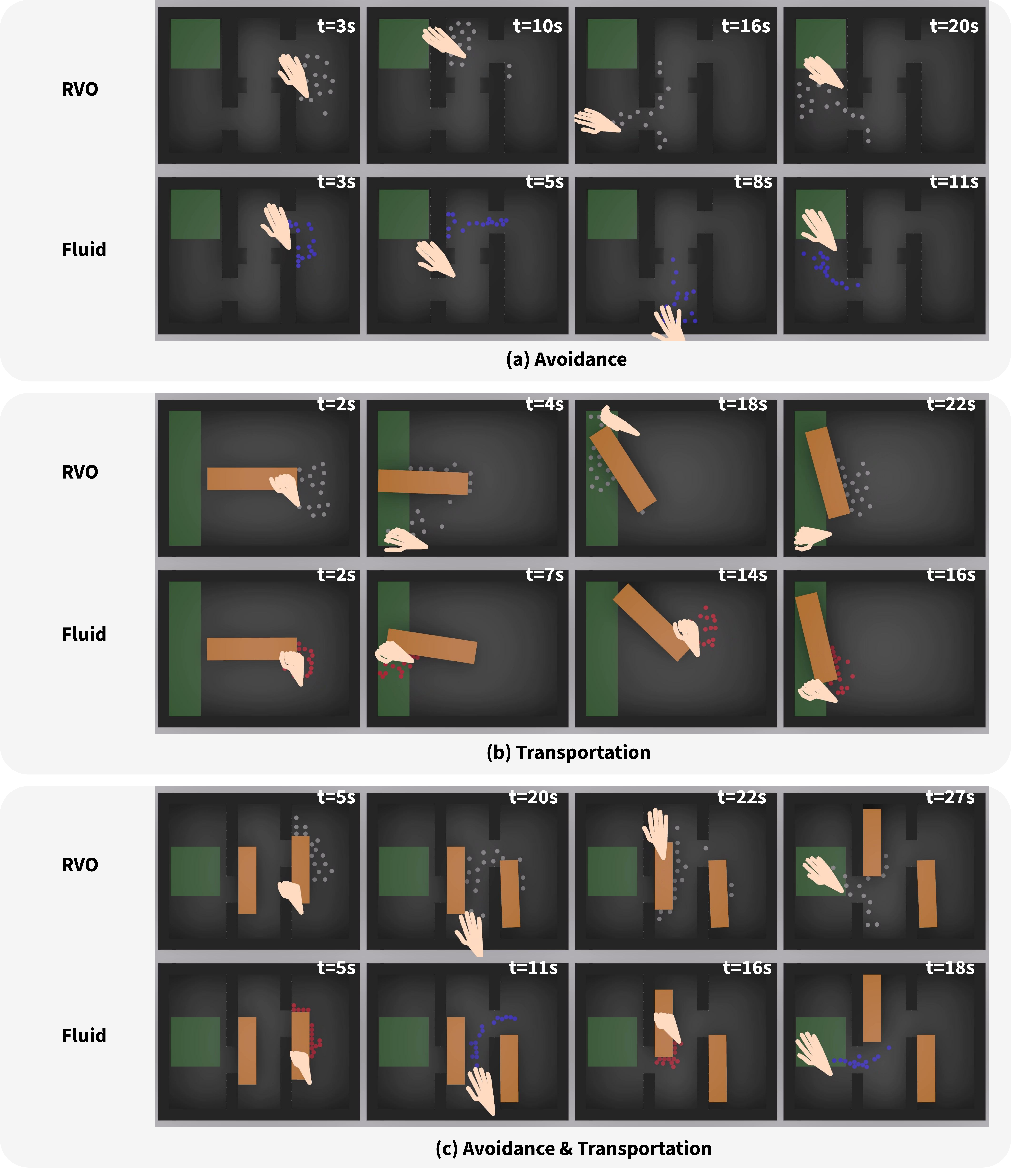}
  \caption{Representative sequences of a participant operating the swarm with the RVO baseline and our fluid-imitation controller (Fluid) in the three environments: (a) avoidance, (b) transportation, and (c) avoidance and transportation. Each panel shows a successful run from a different participant, P2 in (a), P3 in (b), and P16 in (c). With the fluid controller, the swarm robots turn blue in the soft setting and red in the hard setting, so their color shows which formation the participant selected.}
  \Description{Time-lapse screenshots, four per row, showing a participant's hand guiding the swarm with the RVO baseline and our controller across the avoidance, transportation, and avoidance and transportation tasks. With the fluid controller, the swarm robots appear blue in the soft setting and red in the hard setting, indicating which formation the participant selected.}
  \label{fig:user_trajectory}
\end{figure*}

\Cref{fig:user_trajectory} shows how the swarm moved under each controller, and comparing the two controllers task by task reveals where they diverge.

On the avoidance task (\cref{fig:user_trajectory}a), both controllers brought the swarm close to the goal at a similar pace, within roughly the late 10-second range.
The RVO baseline, however, left a few robots stranded behind the obstacles and struggled to gather every robot into the goal.
Our controller reshaped the swarm heavily yet left no robot behind, and the participant kept it in the soft setting throughout.

On the transportation task (\cref{fig:user_trajectory}b), both controllers pushed the object close to the goal at a similar pace.
The RVO swarm, however, spread out around the object and no longer held together as a single mass.
Our controller kept the swarm cohesive, and the participant kept the hard setting throughout.

On the avoidance and transportation task (\cref{fig:user_trajectory}c), the RVO baseline struggled to bring the swarm into the narrow passage and left robots stranded across the environment.
With our controller, the participant switched to the hard setting to push the object out of the way and to the soft setting to move through the narrow passage.
This let them guide the swarm without dropping any robots.

\subsubsection{Semi-Structured Interview}
We first asked participants for their overall impressions of operating the swarm.
Several expected the RVO baseline to work well because its formation looked more hand-like at rest.
After trying the tasks, however, they preferred the fluid controller, especially on the harder avoidance and transportation task.
They described the fluid controller as letting the robots sort out their own arrangement.
They could therefore think less about individual robots and focus on where to go (P8, P11, P13).
With the RVO baseline, they noticed individual robots leaking out of the intended hand shape and catching on obstacles.
Some reported that trying to correct this drift by moving the hand only pushed the robots further out of place (P9, P16, P17).
This mirrors the avoidance and transportation task, where the RVO swarm lost cohesion at the gaps and often failed to move the object in time.

We then asked whether the swarm felt like a part of their hand and whether it followed their hand, and how this changed across tasks and controllers.
Participants reported the strongest sense of agency with the fluid controller and a closed hand (P9, P10, P17).
Several felt the swarm behaved like a separate controller or a pet rather than a part of their body (P1, P12, P13, P15), and one noted that ownership dropped whenever robots were left behind (P17).
Overall, participants reported feeling a strong sense of agency but comparatively little body ownership.

Finally, we asked whether changing the swarm flexibility through the hand shape felt intuitive.
Most participants found this intuitive, since opening the hand to soften the formation and closing it to harden it matched what they expected (12 of 17 participants).

\begin{table*}[t]
  \centering
  \caption{Success rate and completion time for the two controllers in each environment of the user study ($N = 17$). Success rate is given as a percentage with the number of successful trials in parentheses. Completion time is the mean $\pm$ SD over successfully completed trials; timed-out trials are excluded from the time but counted as failures in the success rate.}
  \label{tab:user_eval}
  \begin{tabular}{lcccccc}
    \toprule
    & \multicolumn{2}{c}{Avoidance} & \multicolumn{2}{c}{Transportation} & \multicolumn{2}{c}{Avoidance \& Transportation} \\
    \cmidrule(lr){2-3} \cmidrule(lr){4-5} \cmidrule(lr){6-7}
    Controller & Success (\%) & Time (s) & Success (\%) & Time (s) & Success (\%) & Time (s) \\
    \midrule
    RVO  & 100.0 (17/17) & $25.4 \pm 9.1$  & 94.1 (16/17) & $36.0 \pm 23.0$ & 47.1 (8/17) & $76.9 \pm 34.2$ \\
    \textbf{Fluid (Ours)} & \textbf{100.0 (17/17)} & $\mathbf{18.1} \pm \mathbf{9.9}$  & \textbf{94.1 (16/17)} & $\mathbf{34.7} \pm \mathbf{16.0}$ & \textbf{100.0 (17/17)} & $\mathbf{38.5} \pm \mathbf{19.9}$ \\
    \bottomrule
  \end{tabular}
\end{table*}

\begin{figure*}[t]
  \centering
  \includegraphics[width=\textwidth]{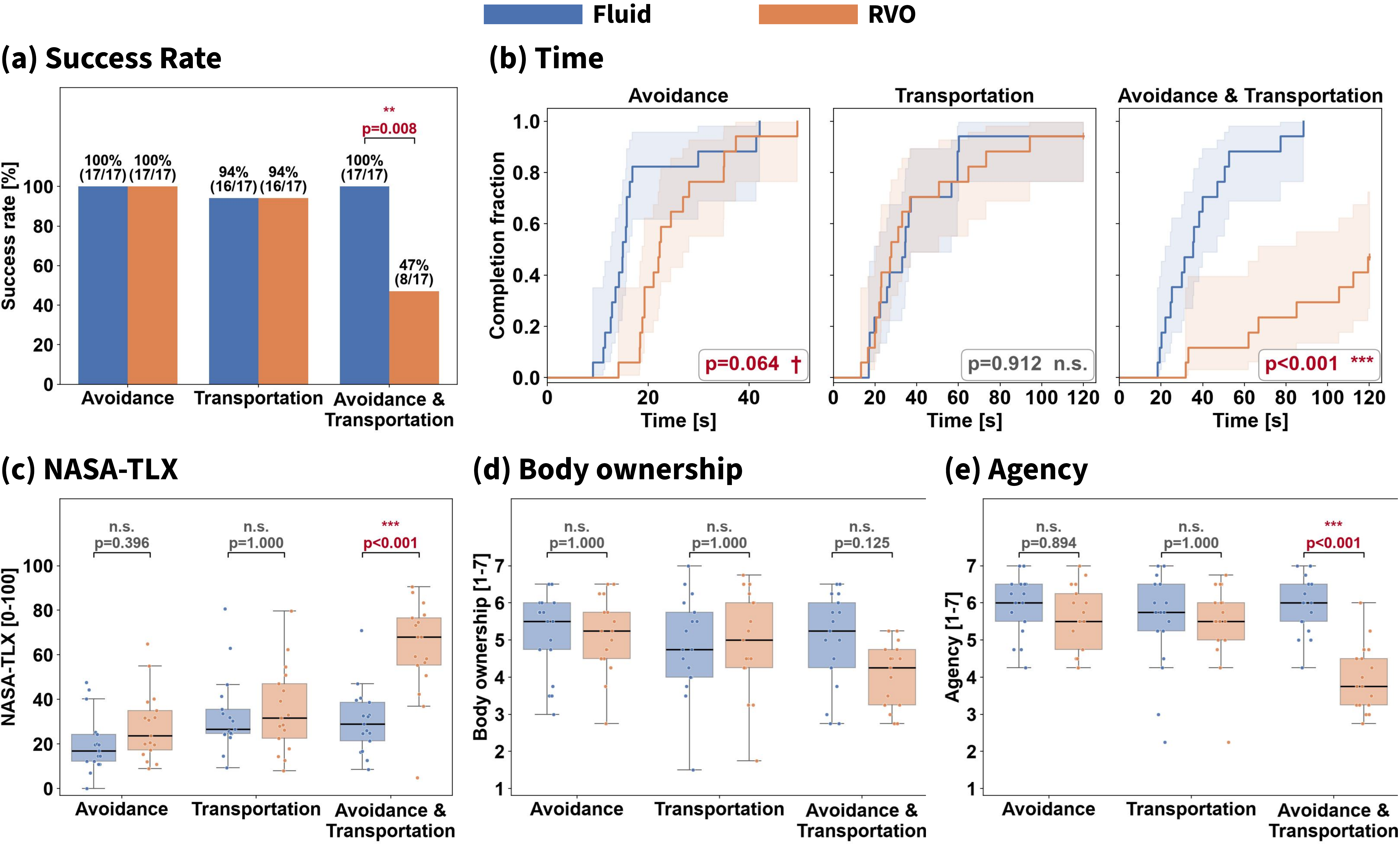}
  \caption{Success rate, completion time, and subjective ratings for the RVO baseline and our controller across the three environments: (a) success rate, (b) completion time, (c) NASA-TLX workload, (d) body ownership, and (e) agency. The two controllers diverge most on the avoidance and transportation task. Significance markers: $^{\dagger}p<.10$, $^{*}p<.05$, $^{**}p<.01$, and $^{***}p<.001$; n.s.\ denotes no significant difference.}
  \Description{Five panels comparing the RVO baseline and our controller across the avoidance, transportation, and avoidance-and-transportation environments: (a) success rate, (b) completion time, (c) NASA-TLX workload, (d) body ownership, and (e) agency. Differences favoring our controller are largest in the avoidance and transportation environment.}
  \label{fig:user_subjective}
\end{figure*}

\section{Real-world Demonstration}
We had participants perform the obstacle avoidance and object transportation tasks to examine whether the experience also holds on physical robots.
We ran this as a public demonstration at ACM SIGGRAPH 2025 Emerging Technologies~\cite{10.1145/3721257.3734028} rather than as a controlled experiment.
It therefore shows that the controller can drive a physical swarm for untrained users and how those users engaged with it.
It does not provide a comparison against a baseline.

This demonstration asks:

\noindent{\itshape\textbf{Q4.} Does the operating experience carry over to a physical swarm?}

\subsection{Apparatus}
\begin{figure}[t]
  \centering
  \includegraphics[width=\linewidth]{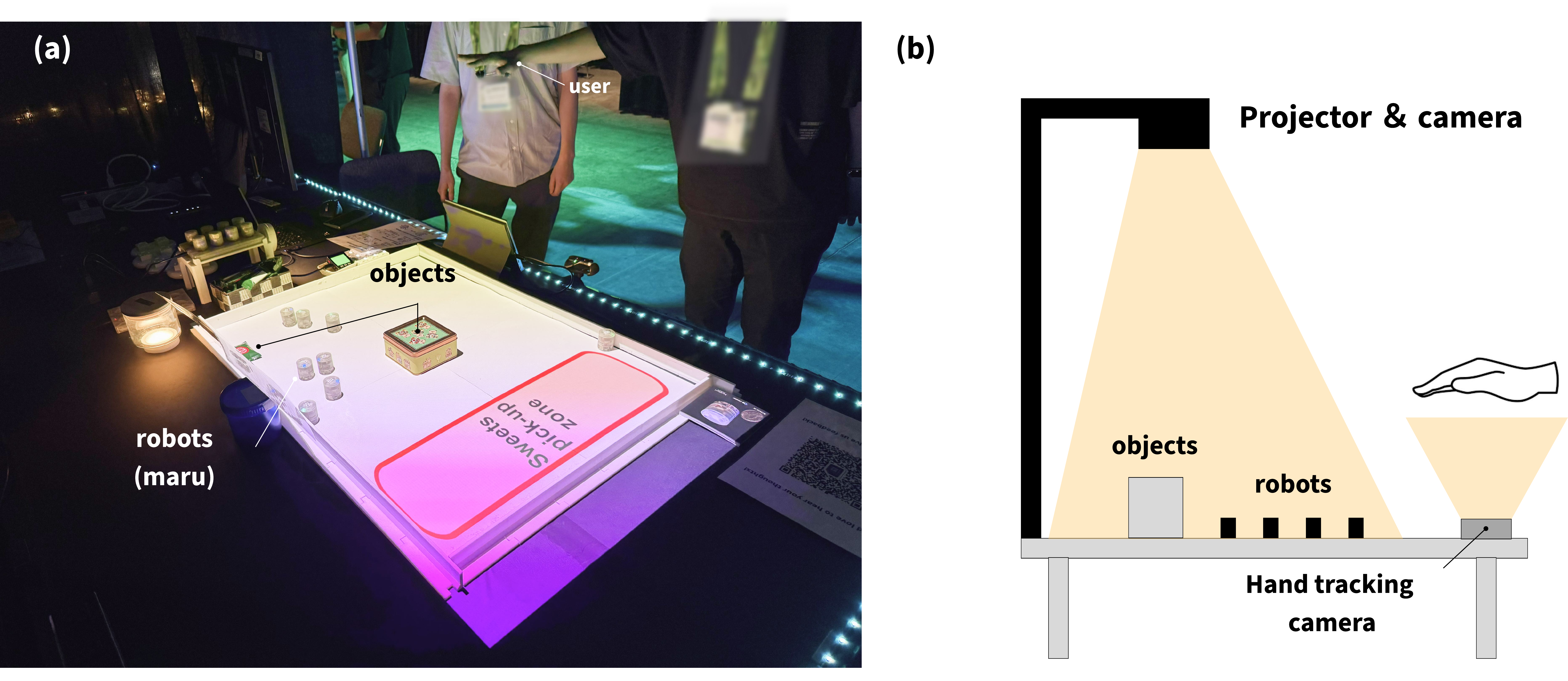}
  \caption{Real-world demonstration apparatus. (a) A photograph of the setup, where the user guides the swarm robots on the table with a hand to avoid objects and transport them to the pick-up zone. (b) A schematic side view: a projector above the table localizes the robots and a camera beside it tracks the objects, while a hand-tracking camera on the table captures the user's hand from below.}
  \Description{Two panels of the real-world demonstration apparatus. (a) A photograph of a table lit from above by the projector, with several small two-wheeled robots, objects to transport, a highlighted pick-up zone, and a user reaching over the table with a bare hand. (b) A schematic side view showing a projector mounted above the table that localizes the robots and a camera beside it that tracks the objects, and a hand-tracking camera placed on the table capturing the user's hand from below.}
  \label{fig:real_world_apparatus}
\end{figure}

\Cref{fig:real_world_apparatus} shows the setup, and we describe the parts relevant to this demonstration below.
We used two-wheeled robots ``maru''\footnote{https://github.com/omron-sinicx/maru}, each 30~mm in diameter and carrying a microcontroller, a radio module, and two photodiodes.
The robots communicate with a base station over a 2.4~GHz ISM-band radio link, and the base station connects to the host computer through USB.
Each robot localizes itself with a projector-based system.
A high-speed projector mounted above the table casts gray-coded patterns onto the table.
Each robot decodes these patterns with its photodiodes to recover its position and orientation and reports them to the host computer.
An Intel RealSense~D435 camera, mounted next to the projector above the table, tracks the positions of the objects.
A hand tracker (Leap Motion Controller~2 from Ultraleap) placed on the table captures the user's hand from below.
The host computer runs our fluid-imitation controller and sends target positions to the robots.
It updates these commands every 100~ms.
The host computer is an ordinary consumer laptop (Lenovo ThinkPad X13 Gen4) with an AMD Ryzen 7 PRO 7840U CPU and 32~GB of RAM.

\subsection{Task}
On the table we scattered everyday objects as obstacles and marked a pick-up zone (\cref{fig:real_world_apparatus}a).
The user guides the swarm with a hand around these objects (obstacle avoidance) and carries one into the pick-up zone (object transportation).
These two tasks mirror those of the simulation study.
The user operates the swarm through our fluid-imitation controller.

\subsection{Participants and Procedure}
We ran this demonstration over the four exhibition days of ACM SIGGRAPH 2025 Emerging Technologies, where participants operated the swarm and then answered a short questionnaire.
Over the run we collected 88 responses.
Each participant guided the swarm with a hand to avoid the objects and carry one into the pick-up zone.
After a facilitator briefly showed how the hand guides the swarm, each participant operated it freely before responding.
The participants were conference attendees who chose to stop at the demonstration.
We ran no baseline condition, and no fixed trial protocol set how long or in what order each person operated the swarm.

\subsection{Measures}
To gather responses from many participants, we used a short questionnaire that each participant could finish quickly, rather than the full protocol of the user study.
Participants rated three single items on a 5-point Likert scale (1~=~strongly disagree, 5~=~strongly agree).
The items covered body ownership, agency, and effectiveness of flexibility control, one item each (\cref{tab:demo_questionnaire}).
We also collected open comments on the experience through an optional free-text field.

\begin{table}[t]
  \centering
  \caption{Items of the real-world demonstration questionnaire. Each construct was rated with a single item on a 5-point Likert scale.}
  \label{tab:demo_questionnaire}
  \begin{tabular}{@{}p{0.28\linewidth}p{0.64\linewidth}@{}}
    \toprule
    Construct & Item \\
    \midrule
    Body ownership
      & Did you feel as if the robot swarm was part of your own body while operating it? \\
    \midrule
    Agency
      & Did you feel that your hand movements directly controlled the robot swarm's behavior? \\
    \midrule
    Effectiveness of flexibility control
      & Did the flexibility of the robot swarm help you control them more effectively? \\
    \bottomrule
  \end{tabular}
\end{table}

\subsection{Results}
\begin{figure}[t]
  \centering
  \includegraphics[width=\linewidth]{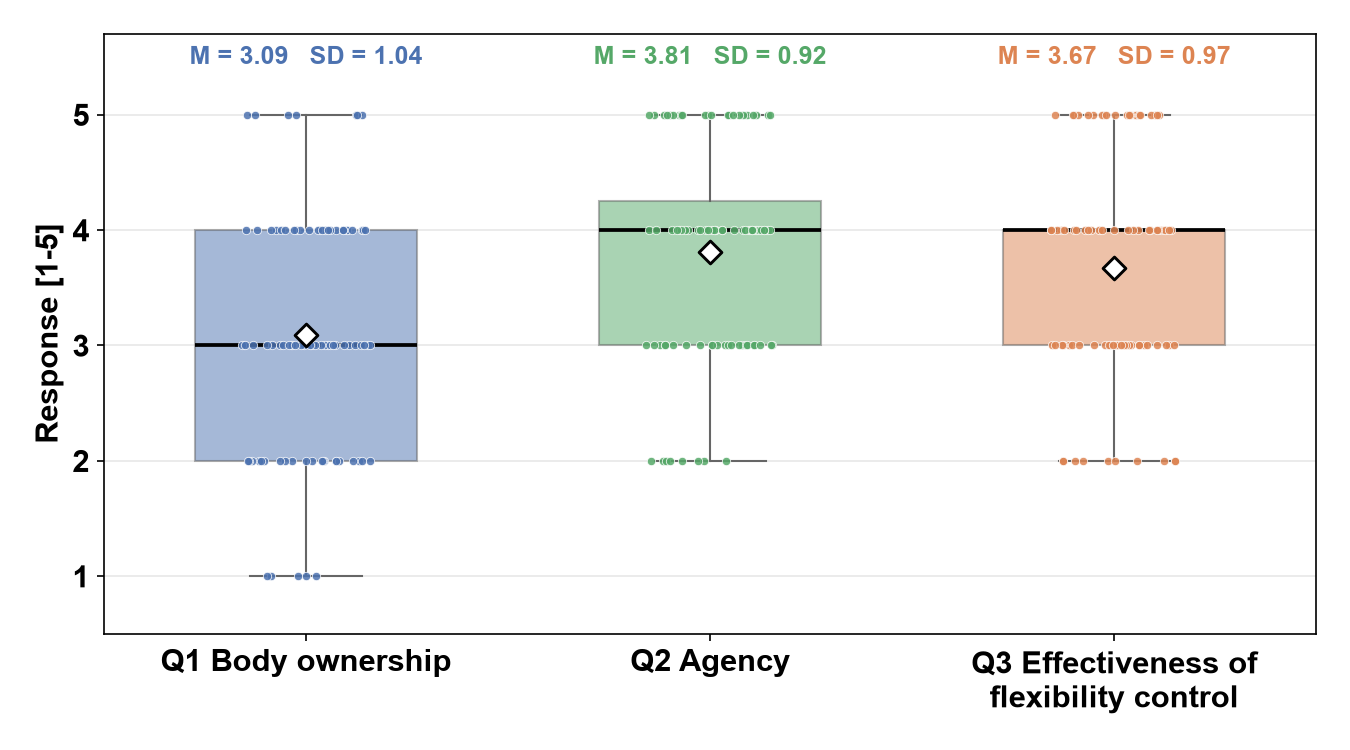}
  \caption{Distribution of the three questionnaire ratings from the real-world demonstration ($N = 88$), on a 5-point Likert scale: body ownership, agency, and effectiveness of flexibility control. Each box shows the median and interquartile range, the diamond marks the mean, and the dots are individual responses; $M$ and $SD$ are printed above each item. Read against the scale midpoint, agency and effectiveness of flexibility control sit clearly above it, while body ownership stays close to it.}
  \Description{Three box plots for the real-world demonstration ratings on a 1-to-5 scale: body ownership (mean 3.09), agency (mean 3.81), and effectiveness of flexibility control (mean 3.67). Each plot overlays individual response dots and a diamond for the mean. Agency and effectiveness of flexibility control sit clearly above the scale midpoint, while body ownership stays close to it.}
  \label{fig:feedback_distribution}
\end{figure}

\Cref{fig:feedback_distribution} shows the distribution of the three ratings.
As an overall trend, all three items were rated at or above the neutral midpoint of the 5-point scale.
Because the three items measure different constructs on separate scales, we read each against that midpoint rather than against one another.
Agency sat furthest above it ($M = 3.81$, $SD = 0.92$, median 4), followed by effectiveness of flexibility control ($M = 3.67$, $SD = 0.97$, median 4).
Body ownership was only slightly above the midpoint ($M = 3.09$, $SD = 1.04$, median 3).

We next turn to the free-text comments we collected from participants.
Many participants described the experience as fun and the swarm as organic or alive.
Some likened it to leading a colony of ants, and several found the control intuitive and easy to learn.
On the other hand, a fair number of comments pointed to weak controllability of the robots.
The most common complaint was latency.
The delay between the hand and the swarm made fine manipulation harder and, for a few participants, weakened the sense of control and ownership.
Others found it hard to sense where the hand sat relative to the tracking camera without looking at a display, and some asked for tighter, more precise following.
We take up these controllability issues in \cref{sec:limitations}.

Overall, participants were able to operate the physical swarm by hand.
Agency and effectiveness of flexibility control were rated clearly above the scale midpoint, while body ownership stayed close to it.
We discuss what this says about transfer to physical robots in \cref{subsec:disc_q4}.

\section{Application}

\begin{figure*}[t]
  \centering
  \includegraphics[width=1.0\linewidth]{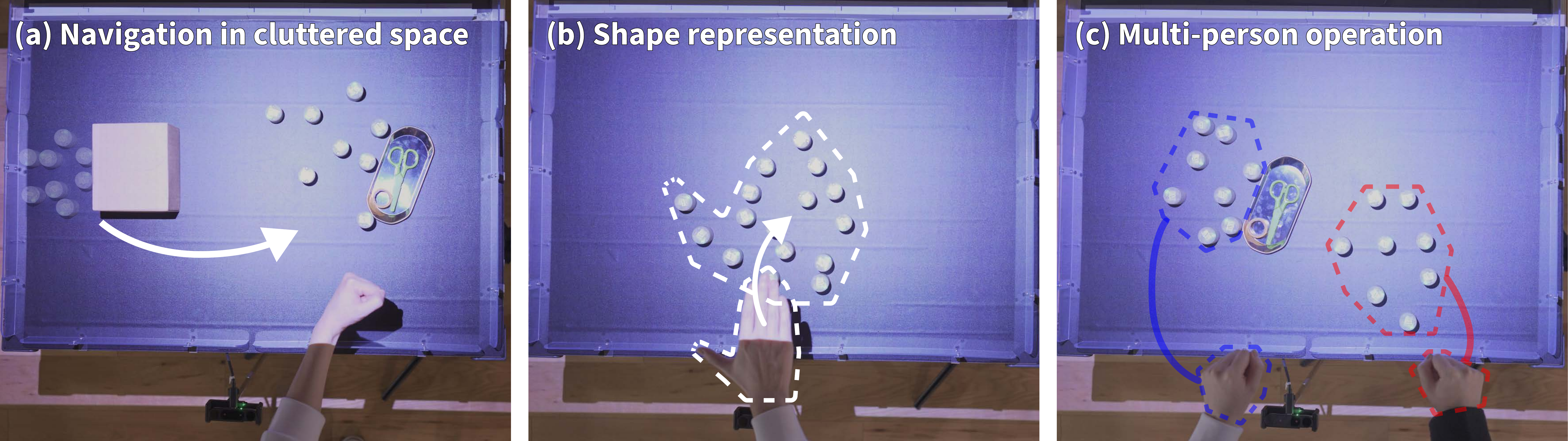}
  \caption{Three applications of Fluidic Body. (a) Navigation in cluttered space: the swarm moves efficiently through a cluttered environment. (b) Shape representation: the swarm takes the shape the user draws with both hands. Two hands allow formations more complex than a single hand can outline. (c) Multi-person operation: two users each drive a swarm of their own at the same time, one outlined in blue and the other in red.}
  \Description{Three photographs of the physical setup taken from above. In (a), a hand guides a group of small white robots past a white box toward a pair of scissors on a tray. In (b), both hands outline a leaf-like shape and the robots arrange themselves along the dashed outline. In (c), two hands operate separate groups of robots, the left one outlined with a blue dashed line and the right one with a red dashed line, with a tray of scissors between them.}
  \label{fig:application}
\end{figure*}

The two tasks we evaluated cover only part of what a hand-driven fluid swarm affords.
We therefore built three applications on the physical setup (\cref{fig:application}).

\begin{itemize}[leftmargin=4mm]
\item \textbf{Navigation in Cluttered Space}: The user guides the swarm across a table strewn with objects and brings a target object back. The soft setting deforms the formation around what stands in the way, and the hard setting closes the robots on the object to move it. One hand thus covers both. This combination is useful in cluttered surroundings such as a lived-in room or a disaster site.

\item \textbf{Shape Representation}: Since the outline of the hand becomes the swarm area, the swarm takes whatever shape the user draws. Two hands extend what is expressible, because an outline that spans both of them is one that a single hand cannot hold.

\item \textbf{Multi-person Operation}: Each swarm follows the hand that shapes it. Two users can therefore drive one swarm each at the same time. This makes possible work that one person alone cannot keep track of, such as operating two formations at once.
\end{itemize}

\section{Discussion}

\subsection{Q1: Task Adaptivity}
\label{subsec:disc_q1}

Matching the fluid properties to each task gives Q1 a positive answer.
One controller reached the best result on each task, whereas fixed parameters fit only one side of the trade-off.
Below, we discuss the successes of each setting and the shortfall of the baseline, and then the limits of this evaluation.

The two fluid settings succeeded and failed in opposite places (\cref{tab:tech_eval}, \cref{fig:tech_eval_results}).
The soft setting cleared avoidance significantly more often than either other controller.
On transportation, the hard setting cleared every trial as RVO did and completed significantly faster than it.
The two parameters explain this split.
The soft setting keeps both the viscosity coefficient and the reference density low.
Neighboring robots therefore move at different velocities and settle farther apart.
This lets the formation deform around an obstacle and slip through a gap.
The same looseness, however, costs the swarm on transportation.
The robots spread around the object more widely than the task needs, and some of them push it from directions the user never intended.
A simple hand motion then no longer drives the object toward the goal.
The hard setting raises both parameters, which aligns the velocities and packs the robots closer together.
The swarm thus stays one mass and drives the object along a single direction.
It fails in its own way on avoidance.
A robot that slips into a small recess of an obstacle cannot work itself free, and the rest of the formation moves on and leaves it behind.

RVO sits on the same side of this split as the hard setting.
It completed every transportation trial but few avoidance trials.
On transportation it follows the commanded pattern straightforwardly, which is enough to push the object into the goal in every trial.
Unlike the hard setting, though, it never adapts the formation to the object as the object moves.
The hard setting accordingly completed the task significantly faster.
On avoidance the same mechanism turns against it.
RVO gives each robot a velocity that clears the obstacles and neighbors around it while heading to the goal.
A robot deflected by an obstacle is thus deflected on its own.
RVO has no counterpart to the fluid force that draws a stray robot back toward the group.
Consequently, even a few robots left behind an obstacle were enough to fail a trial.
The same failure surfaced again once participants drove RVO themselves (\cref{fig:user_trajectory}a).
As a result, RVO falls significantly behind a fluid setting on both tasks.

Because the soft formation deforms around obstacles, its success on avoidance could be credited to deformability alone.
Yet even the hard setting, the stiffer of the two fluid settings, cleared avoidance significantly more often than RVO.
Part of the fluid settings' advantage therefore comes from the fluid force that ties each robot to its neighbors, not only from the deformability of the formation.

Overall, neither fluid setting alone covered both tasks, since the soft setting failed on transportation and the hard setting fell short on avoidance.
Covering both tasks therefore requires the user to switch between the two settings as each task demands.
This evaluation fixed one setting per condition in advance.
The user study thus tests whether a person can make this switch while operating the swarm.

\subsection{Q2: Task Efficiency}
\label{subsec:disc_q2}

Q2 also has a positive answer, and our controller's advantage grows as the task gets harder.
Only on the combined task did participants both clear every trial, where the baseline often ran out of time, and finish significantly faster.
On transportation, the two controllers did not differ significantly, and on avoidance the difference was only marginally significant.
The rest of this section traces the source of each difference and its bearing on the adaptivity established in \cref{subsec:disc_q1}.

The avoidance and transportation task is the only one that asks for both demands inside a single trial, and our controller outperformed the baseline there most clearly (\cref{tab:user_eval}).
Participants switched to the hard setting to push the object clear of the passage and to the soft setting to bring the swarm through it (\cref{fig:user_trajectory}c).
One run thus met both requirements in turn.
The baseline could not trade one demand for the other.
Its swarm lost cohesion at the gap and left robots scattered across the field, and the trials that ran out of time ended that way.
Participants reported the same thing in the interview.
They said the fluid controller let the robots settle their own arrangement, which freed them to attend to where the swarm was going rather than to individual robots.

On the two single-demand tasks both controllers satisfied what the task asked.
With either controller, participants cleared almost every trial.
On the avoidance task, our controller still tended to be faster, and the difference was marginally significant.
With the baseline, robots stayed behind the obstacles, and participants had to spend the remaining seconds gathering them into the frame.
The soft formation instead deformed past the obstacles and arrived together (\cref{fig:user_trajectory}a).
On the transportation task, we found no performance difference, since a fixed formation can meet every demand of pushing an object across open space.
The technical evaluation agrees on reliability, where the hard setting and the baseline cleared the task equally often.
It did show the hard setting completing significantly faster there.
Once a person drove the swarm, the spread between participants absorbed this margin.
Taken together, how much this matters depends on how hard the task is.
A fixed formation is enough for a single demand, where adjusting the fluid parameters only saves time.
In contrast, when two demands meet in one trial, adjusting the fluid parameters decides whether the task succeeds.

Participants also chose the same settings that the technical evaluation found best for each task.
They held soft through avoidance and hard through transportation (\cref{fig:user_trajectory}a, b), and switched between them only when a trial demanded both.
We showed them how to switch between soft and hard but never how to solve a task.
Each participant therefore judged on their own which setting suited the task in front of them.

\subsection{Q3: Operating Experience}
\label{subsec:disc_q3}

Q3 has a positive answer, and it comes from the same task where efficiency improved.
Workload and agency both favored our controller significantly on the combined task and showed no controller difference on the two simpler ones.
Body ownership moved the same way without reaching significance.
The paragraphs below cover the joint change in experience and efficiency, the interview evidence linking the two, and the difference between agency and ownership.

Workload separated the two controllers only on the combined task, where the baseline's roughly doubled while ours held a similar value across all three (\cref{fig:user_subjective}c).
The interviews suggest how the load and the efficiency are connected.
Participants said the fluid controller let the robots settle their own arrangement.
They could then think about where the swarm was going rather than about individual robots (P8, P11, P13).
With the baseline they watched robots leak out of the intended hand shape and catch on obstacles.
Several participants found that correcting this drift by moving the hand only pushed the robots further out of place (P9, P16, P17).
Attention spent recovering stragglers is attention taken from the task itself, which fits both the heavier workload and the timeouts the baseline produced on the combined task.

Agency behaved exactly as workload did.
It held a similar value across all three tasks with our controller, while the baseline fell away significantly on the combined one (\cref{fig:user_subjective}e).
Body ownership pointed the same way but more weakly (\cref{fig:user_subjective}d).
It favored our controller most on the combined task, yet on no task did the difference between the two controllers reach significance.
Agency also scored above ownership on every task.
The interviews are consistent with this gap.
Participants described the swarm as a separate controller or a pet rather than as a part of their body (P1, P12, P13, P15).
One reported that ownership dropped whenever robots were left behind (P17).
This split reflects the division of labor between the hand and the robots.
The hand sets where the swarm goes and how it behaves, and the swarm does exactly that.
The user therefore feels in command of it.
The individual robots, however, take their own positions along the way, which the user's own fingers never do.
The swarm thus stays something the user operates rather than a part of the user's own body.

Read alongside \cref{subsec:disc_q2}, the baseline's losses in efficiency and in experience share one cause, the robots it left behind at the gaps.
Once robots were lost, the swarm stopped following the hand, and participants registered this as a loss of agency.
One of them tied ownership directly to robots being left behind (P17).
We believe that understanding how the swarm moves lets a person operate it efficiently, and the success rates and completion times are consistent with this view.
Overall, participants changed the swarm's behavior through the fluid properties as they intended.
This kept the swarm following the hand even on the combined task, where holding its behavior together is hardest.
The operating experience held there rather than degrading, and we read this as the reason the efficiency held as well.

\subsection{Q4: Transfer to Physical Robots}
\label{subsec:disc_q4}
Q4 has a positive answer.
Because the three constructs were each measured on their own single-item scale, we read each against the neutral midpoint rather than against one another. Participants guided the physical swarm through similar avoidance and transportation tasks.
They placed agency and effectiveness of flexibility control clearly above the midpoint, while body ownership stayed close to it. This echoes the VR user study, where agency likewise ran ahead of body ownership.

Responsiveness, on the other hand, is where the physical setting fell behind.
The delay between the hand and the swarm was the most common complaint in the free-text responses.
A few participants tied it to a weaker sense of control and ownership, and we take up this issue in \cref{sec:limitations}.

Overall, moving to physical robots cost the experience some of its quality, mostly through the delay between the hand and the swarm.
The direction of that experience, however, carried over.

\subsection{Generality of the Interaction Paradigm}
\label{subsec:disc_generality}

The key element of our approach is the fluid layer between the hand and the swarm.
Hand input often lacks the dimensions to control a whole swarm, but with the fluid layer the user only sets a material property, and the robots compute their own states.

Although all our evaluations used mobile robots on a surface, the approach applies in principle to any system of many actuated elements whose individual states a person cannot practically specify.
Shape displays, actuated tangibles, and robots serving as physical pixels~\cite{Le_Goc2016-dg, Suzuki2019-ey, Hiraki2018-gi} could also expose a material property instead of the state of each element.

The input is also not limited to a single person's hand, and other body parts or other people's hands could shape the fluid as well.
Each user controls their own fluid, and several swarms can therefore share one table.
A swarm could also pass from one user to another when a different user outlines it, with no device changing hands.

\section{Limitations and Future Work}
\label{sec:limitations}

\subsection{Latency and Responsiveness}
The delay between the hand and the swarm was the most common complaint in the physical demonstration.
It comes from the implementation rather than the controller, since the same controller drew no such complaint in VR.
Two factors in the implementation add to this delay.
First, the radio link lets the host send commands only every 100~ms.
Second, the robots drive on two wheels and have to turn before they move.
Because neither factor is easy to remove, predicting the hand's motion between updates, or compensating for the delay elsewhere in the loop, would be more effective than removing it.

\subsection{Scope of Fluid Parameters}
The user controls two fluid properties, the viscosity and the reference density.
The swarm therefore changes both its flexibility and its spread.
Exposing more of the fluid's properties, and controlling them continuously, would give the user a wider range of control.
Our evaluations, however, compared only the two discrete settings, soft and hard, and testing the swarm's behavior between and beyond them is left for future work.

\subsection{Hand-to-Parameter Mapping}
Hand movements as user input and fluid properties as control parameters can be paired in many ways.
Our mapping is one design among them, and we fixed it provisionally.
We asked about the mapping in only a single interview question, and most participants called opening and closing the hand an intuitive way to change the formation's flexibility.
The free-text responses from the physical demonstration pointed the same way.


\subsection{Interaction with a Three-Dimensional Swarm}
This work stays on a tabletop plane, since our robots drive on two wheels and localize themselves from a projector above the table.
As for the controller, SPH is formulated in three dimensions and we reduced it to two for our setup, which makes extending it to a three-dimensional swarm straightforward.
However, operating such a swarm is harder, since the table currently fixes one dimension and the hand only has to specify the other two.
A swarm that hovers overhead or surrounds the body leaves all three to the user, who must also set how far the fluid reaches.
The experience would change as well, since the swarm would surround the user instead of sitting in front of them.
Agency and body ownership in that setting would be an interesting question for future work.

\subsection{Scope of the Controller Comparison}
Both evaluations compared our controller with a single baseline, RVO.
However, relying on this one baseline leaves two points open.

First, the baseline kept its parameters fixed from start to finish in every trial.
We kept RVO fixed because we wanted to compare our approach with existing swarm robot interfaces, which use RVO this way and let their users direct only the robots' positions.
Part of the advantage we measured may therefore come from adjustability itself rather than from the fluid.
As reviewed in the related work, some swarm controllers do change their behavior with the task~\cite{engebraaten2020multifunction, batra2022decentralized, chen2024stgnn}.
This work, however, targets behavior adaptation by a human operator, and few controllers offer parameters that a person can change during operation.
We therefore used RVO as the only baseline.

Second, the parameters we used may not be optimal.
We chose the two fluid settings through a search in pilot trials and ran RVO with the standard parameters of its original formulation~\cite{Van_den_Berg2008-we}.
However, no theory guarantees that either choice is the best.
For example, RVO with a shorter time horizon or a smaller neighbor distance might pass the gaps that stranded it on the avoidance task.
However, such tuning of RVO might improve avoidance at the cost of stability in transportation.
More carefully designed fluid parameters might also lead to different behavior on both tasks.

\subsection{Operating Load as the Swarm Scales}
We used 16 robots in the technical evaluation and kept the number fixed in every evaluation.
Operating a swarm should get harder as the swarm grows, since a larger group gives the user more to keep track of.
Our controller should suffer less from this, because the user addresses the fluid rather than the robots.
The hand therefore does the same work however many robots the swarm contains, although the present work does not show this.
Measuring the change in operating load with the number of robots would require a larger fleet than we had, but the question is worth pursuing.


\section{Conclusion}
We proposed Fluidic Body, which places a virtual fluid between the hand and the swarm so that a person shapes the swarm's behavior, not only its destination.
Reaching the swarm through a fluid lets a person guide the robots by intuition rather than by instruction, and adjust their movement as each task requires.

Three evaluations support this design.
In a simulation study run without a human operator, each fluid setting suited a different task, and our controller cleared both obstacle avoidance and object transportation. An existing method that behaves the same way in every task managed only one of the two.
In a VR user study, 17 participants drove the swarm themselves and completed the combined avoidance-and-transportation task faster and more reliably than with that method. They also reported a lower workload and a stronger sense of agency.
In a real-world demonstration, 88 participants operated physical robots. Their responses followed the same trend, and they agreed most strongly that their own hand movements drove the swarm.

Fluidic Body shows that a swarm becomes easier to operate once the person reaches it through a fluid, something whose behavior they already know from everyday life.
Working through the fluid also shortens the distance between the person and the swarm, and gives them what feels like a body whose properties change to suit the task.

\begin{acks}
This work was supported by JSPS KAKENHI Grant Number JP25K22837.
\end{acks}

\bibliographystyle{ACM-Reference-Format}
\bibliography{fluidic_swarm}



\end{document}